\documentclass[10pt,twocolumn,journal]{IEEEtran}
\usepackage{setspace}
\usepackage{clrscode}
\usepackage{pict2e}

\usepackage{amsmath}
\usepackage{amssymb}
\usepackage[dvips]{graphicx}
\usepackage{epsfig}
\usepackage{caption}
\usepackage{subcaption}

\newcommand{\rme}{{\mathrm{e}}}

\newcommand{\E}{\mathbb{E}}

\newcommand{\mH}{\mathcal{H}}
\newcommand{\hH}{\hat{\mathcal{H}}}

\makeatletter

\newcommand{\Rmnum}[1]{\expandafter\@slowromancap\romannumeral #1@}
\makeatother

\newcommand{\by}{\bar{y}}
\newcommand{\byr}{\bar{y}_r}
\newcommand{\byi}{\bar{y}_i}
\newcommand{\bn}{\bar{n}}
\newcommand{\bnr}{\bar{n}_r}
\newcommand{\bni}{\bar{n}_i}
\newcommand{\bw}{\bar{w}}
\newcommand{\bwr}{\bar{w}_r}
\newcommand{\bwi}{\bar{w}_i}

\newcommand{\argmax}{\operatornamewithlimits{arg\,max}}

\newcommand{\SEP}{\text{SEP}}

\newcommand{\avg}{\text{avg}}

\newcommand{\nid}{\rm{d}}
\newcommand{\f}{\rm{f}}
\newcommand{\pk}{\rm{pk}}

\newtheorem{Rem}{Remark}

\newtheorem{Prop}{Proposition}

\begin{document}

\title{Error Rate Analysis of Cognitive Radio Transmissions with Imperfect Channel Sensing}

\author{\vspace{1cm}
\authorblockN{Gozde Ozcan, M. Cenk Gursoy, and Sinan Gezici}
\thanks{Gozde Ozcan and M. Cenk Gursoy are with the Department of Electrical
Engineering and Computer Science, Syracuse University, Syracuse, NY, 13244
(e-mail: gozcan@syr.edu, mcgursoy@syr.edu).}
\thanks{Sinan Gezici is with the Department of Electrical and Electronics Engineering, Bilkent University, Bilkent, Ankara 06800, Turkey (e-mail: gezici@ee.bilkent.edu.tr).}
}

\maketitle

\begin{abstract}
This paper studies the symbol error rate performance of cognitive radio transmissions in the presence of imperfect sensing decisions. Two different transmission schemes, namely sensing-based spectrum sharing (SSS) and opportunistic spectrum access (OSA), are considered. In both schemes, secondary users first perform channel sensing, albeit with possible errors. In SSS, depending on the sensing decisions, they adapt the transmission power level and coexist with primary users in the channel. On the other hand, in OSA, secondary users are allowed to transmit only when the primary user activity is not detected. Initially, for both transmission schemes, general formulations for the optimal decision rule and error probabilities are provided for arbitrary modulation schemes under the assumptions that the receiver is equipped with the sensing decision and perfect knowledge of the channel fading, and the primary user's received faded signals at the secondary receiver has a Gaussian mixture distribution. Subsequently, the general approach is specialized to rectangular quadrature amplitude modulation (QAM). More specifically, the optimal decision rule is characterized for rectangular QAM, and closed-form expressions for the average symbol error probability attained with the optimal detector are derived under both transmit power and interference constraints. The effects of imperfect channel sensing decisions, interference from the primary user and its Gaussian mixture model, and the transmit power and interference constraints on the error rate performance of cognitive transmissions are analyzed.
\end{abstract}

\thispagestyle{empty}


\section{Introduction}

Rapid growth in the use of wireless services coupled with inefficient utilization of scarce spectrum resources has led to much interest in the analysis and development of cognitive radio systems. Hence, performance analysis of cognitive radio systems is conducted in numerous studies to gain more insights into their potential uses. In most of the previous work, transmission rate is considered as the main performance metric. For instance, the secondary user mean capacity was studied in \cite{smith} by imposing a constraint on the signal-to-interference-noise ratio (SINR) of the primary receiver and considering different channel side information (CSI) levels. The authors in \cite{kang} determined the optimal power allocation strategies that achieve the ergodic capacity and the outage capacity of the cognitive radio channel under various power and interference constraints. In \cite{stotas}, the authors studied the optimal sensing time and power allocation strategy to maximize the average throughput in a multiband cognitive radio network. Recently, the work in \cite{rezki} proposed generic expressions for the optimal power allocation scheme and the ergodic capacity of a spectrum sharing cognitive radio under different levels of knowledge on the channel between the secondary transmitter and the secondary receiver and the channel between the secondary transmitter and the primary receiver subject to average/peak transmit power constraints and the interference outage constraint.

Although transmission rate is a common performance metric considered for secondary users, error rate is another key performance measure to quantify the reliability of cognitive radio transmissions. In this regard, several recent studies incorporate error rates in cognitive radio analysis \cite{asghari}--\cite{Li}. For instance, the authors in \cite{asghari} characterized the optimal constellation size of $M$-QAM and the optimal power allocation scheme that maximize the channel capacity of secondary users for a given target bit error rate (BER), interference and peak power constraints. The work in \cite{Islam} mainly focused on the power allocation scheme minimizing the upper bound on the symbol error probability of phase shift keying (PSK) in multiple antenna transmissions of secondary users. The authors in \cite{Do} proposed a channel switching algorithm for secondary users by exploiting the multichannel diversity to maximize the received SNR at the secondary receiver and evaluated the transmission performance in terms of average symbol error probability. The optimal antenna selection that minimizes the symbol error probability in underlay cognitive radio systems was investigated in \cite{Sarvendranath}. Moreover, the recent work in \cite{wang} analyzed the minimum BER of a cognitive transmission subject to both average transmit power and interference power constraints. In their model, the secondary transmitter is equipped with multiple antennas among which only one antenna that maximizes the weighted difference between the channel gains of transmission link from the secondary transmitter to the secondary receiver and interference link from the secondary transmitter to the primary receiver is selected for transmission. The authors in \cite{Suraweera} obtained a closed-form BER expression under the assumption that the interference limit of the primary receiver is very high. Also, the work in \cite{Xu} focused on the optimal power allocation that minimizes the average BER subject to peak/average transmit power and peak/average interference power constraints while the interference on the secondary users caused by primary users is omitted.  Moreover, in \cite{Li}, the opportunistic scheduling in multiuser underlay cognitive radio systems was studied in terms of link reliability.

In the error rate analysis of the above-mentioned studies, channel sensing errors are not taken into consideration. Practical cognitive radio systems, which employ spectrum sensing mechanisms to learn the channel occupancy by primary users, generally operate under sensing uncertainty arising due to false alarms and miss-detections.
For instance, different spectrum sensing methods for Gaussian \cite{Ghasemi}, \cite{Axell} and non-Gaussian environments \cite{Moghimi}, \cite{Zahabi} and dynamic spectrum access strategies \cite{Hossain} have extensively been studied recently in the literature, and as common to all schemes, channel sensing is generally performed with errors and such errors can lead to degradation in the performance.

With this motivation, we in this paper study the symbol error rate performance of cognitive radio transmissions in the presence of imperfect channel sensing decisions. We assume that secondary users first sense the channel in order to detect the primary user activity before initiating their own transmissions. Following channel sensing, secondary users employ two different transmission schemes depending on how they access the licensed channel: sensing-based spectrum sharing (SSS) and opportunistic spectrum access (OSA). In the SSS scheme \cite{Kang2}, cognitive users are allowed to coexist with primary users in the channel as long as they control the interference by adapting the transmission power according to the channel sensing results. More specifically, secondary users transmit at two different power levels depending on whether the channel is detected as busy or idle. In the OSA scheme \cite{Akyildiz}, cognitive users are allowed to transmit data only when the channel is detected as idle, and hence secondary users exploit only the silent periods in the transmissions of primary users, called as spectrum opportunities. Due to the assumption of imperfect channel sensing, two types of sensing errors, namely false alarms and miss detections, are experienced. False alarms result in inefficient utilization of the idle channel while miss-detections lead to cognitive users' transmission interfering with primary user's signal. Such interference can be limited by imposing interference power constraints.


In our error rate analysis, we initially formulate the optimal decision rule and error rates for an arbitrary digital modulation scheme. Subsequently, motivated by the requirements to efficiently use the limited spectrum in cognitive radio settings, we concentrate on quadrature amplitude modulation (QAM) as it is a bandwidth-efficient modulation format. More specifically, in our analysis, we assume that the cognitive users employ rectangular QAM for data transmission, analysis of which, as another benefit, can easily be specialized to obtain results for square QAM, pulse amplitude modulation (PAM), quadrature phase-shift keying (QPSK), and binary phase-shift keying (BPSK) signaling.


In addition to the consideration of sensing errors and relatively general modulation formats, another contribution of this work is the adoption of a Gaussian mixture model for the primary user's received faded signals in the error-rate analysis. The closed-form error rate expressions in aforementioned works \cite{asghari}--\cite{Li} are obtained when primary user's faded signal at the secondary receiver is assumed to be Gaussian distributed. However, in practice, cognitive radio transmissions can be impaired by different types of non-Gaussian noise and interference, e.g., man-made impulsive noise \cite{Middleton}, narrowband interference caused by the primary user \cite{Giorgetti}, primary user's modulated signal \cite{liang}, and co-channel interference from other cognitive radios \cite{Hu}.  Therefore, it is of interest to investigate the error rate performance of cognitive radio transmissions in the presence of primary user's interference which is modeled to have a Gaussian mixture probability density function (pdf) (which includes pure Gaussian distribution as a special case) \cite{Anhari}.

Main contributions of this paper can be summarized as follows. Under the above-mentioned assumptions, we first derive, for both SSS and OSA schemes, the optimal detector structure, and then we present a closed-form expression of the average symbol error probability under constraints on the transmit power and interference. Through this analysis, we investigate the impact of imperfect channel sensing (i.e., the probabilities of detection and false alarm), interference from the primary user, and both transmit power and interference constraints on the error rate performance of cognitive transmissions. Also, the performances of SSS and OSA transmission schemes are compared when primary user's faded signal is modeled to have either a Gaussian mixture or a purely Gaussian density.

The remainder of this paper is organized as follows: Section \ref{sec:system_model} introduces the system model. In Section \ref{sec:general_formulation}, general formulations for the optimal detection rule and average symbol error probability in the presence of channel sensing errors are provided for SSS and OSA schemes. In Section \ref{sec:error_rate_QAM_PAM}, closed-form average symbol error probability expressions for specific modulation types, i.e., arbitrary rectangular QAM and PAM are derived subject to both transmit power and interference constraints under the assumptions of Gaussian-mixture-distributed primary user faded signal and imperfect channel sensing decisions. Numerical results are provided and discussed in Section \ref{sec:num_results}. Finally, Section \ref{sec:conclusion} concludes the paper.

\section{System Model} \label{sec:system_model}

\subsection{Channel Sensing} \label{subsec:sensing}

We consider a cognitive radio system consisting of a pair of secondary transmitter-receiver and a pair of primary transmitter-receiver\footnote{As noted in the subsequent subsections, the analysis in the paper can be extended to account for more than one primary transmitter-receiver pair by 1) slightly modifying the interference constraints to limit the worst-case interference on multiple primary receivers and 2) selecting a Gaussian mixture density that reflects the distribution of the received faded sum signal of multiple primary transmitters.}. The secondary user initially performs channel sensing, which can be modeled as a hypothesis testing problem. Assume that $\mH_0$ denotes the hypothesis that the primary users are inactive in the channel, and $\mH_1$ denotes the hypothesis that the primary users are active. Various channel sensing methods, including energy detection, cyclostationary detection, and matched filtering, have been proposed and analyzed in the literature. Regardless of which method is used, one common feature is that errors in the form of miss-detections and false-alarms occur in channel sensing. The ensuing analysis takes such errors into account and depends on the sensing scheme only through the detection and false-alarm probabilities. Assume that $\hH_0$ and $\hH_1$ denote the sensing decisions that the primary users are inactive and active, respectively. Then, the detection and false-alarm probabilities can be expressed respectively as the following conditional probabilities:
\begin{align}
P_{\nid} &= \Pr\{\hH_1 | \mH_1\},\\
P_{\f} &= \Pr\{\hH_1 | \mH_0\}.
\end{align}

\subsection{Power and Interference Constraints} \label{subsec:power-interference}

Following channel sensing, the secondary transmitter performs data transmission over a flat-fading channel. In the SSS scheme, the average transmission power is selected depending on the channel sensing decision. More specifically, the average transmission power is $P_1$ if primary user activity is detected in the channel (denoted by the event $\hH_1$) whereas the average power is $P_0$ if no primary user transmissions are sensed (denoted by the event $\hH_0$). We assume that there is a peak constraint on these average power levels, i.e., we have
\begin{align}
\label{eq:peak_power_constraint}
P_0 \le P_{\pk} \quad \text{ and } \quad P_1 \le P_{\pk},
\end{align}
where $P_{\pk}$ denotes the peak transmit power limit. We further impose an average interference constraint in the following form:
\begin{align}
\label{eq:interference_power_constraint_SSS}
(1-P_{\nid})\,P_0 \,\E\{|g|^2\} + P_{\nid} \,P_1 \, \E\{|g|^2\} \le Q_{\avg}
\end{align}
where $P_{\nid}$ is the detection probability and $g$ is the channel fading coefficient between the secondary transmitter and the primary receiver. Note that with probability $P_{\nid}$, primary user activity is correctly detected and primary receiver experiences average interference proportional to $P_1 \E\{|g|^2\}$. On the other hand, with probability $(1-P_{\nid})$, miss-detections occur, secondary user transmits with power $P_0$, and primary receiver experiences average interference proportional to $P_0 \E\{|g|^2\}$. Therefore, $Q_{\avg}$ can be regarded as a constraint on the average interference inflicted on the primary user\footnote{Note that the rest of the analysis can easily be extended to the case of $M$ primary receivers by replacing the constraint in (\ref{eq:interference_power_constraint_SSS}) with $(1-P_{\nid})P_0 \max_{1\le i \le M} \E\{|g_i|^2\} + P_{\nid} P_1 \max_{1\le i \le M} \E\{|g_i|^2\} \le Q_{\avg}$, where $g_i$ is the channel fading coefficient between the secondary transmitter and the $i^{\text{th}}$ primary receiver. In this setting, $Q_{\avg}$ effectively becomes a constraint on the worst-case average interference.}.

In the OSA scheme, no transmission is allowed when the channel is detected as busy and hence, we set $P_1$ = 0. Now with the peak power and average interference constraints, we have
\begin{align}
P_0 \le P_{\pk} \quad \text{ and } \quad (1-P_{\nid})\,P_0 \,\E\{|g|^2\} \le Q_{\avg}
\end{align}
which can be combined to write
\begin{align}
\label{eq:interference_power_constraint_OSA}
P_0 \le \min\left\{ P_{\pk}, \frac{Q_{\avg}}{(1-P_{\nid}) \E\{|g|^2\}}\right\}.
\end{align}

Above, we have introduced the average interference constraint. However, if the instantaneous value of the fading coefficient $g$ is known at the secondary transmitter, then a peak interference constraint in the form
\begin{align}
P_i |g|^2 \le Q_{\pk}
\end{align}
for $i = 0,1$ can be imposed. Note that transmission power $P_0$ in an idle-sensed channel is also required to satisfy the interference constraint due to sensing uncertainty (i.e., due to the consideration of miss-detection events). Hence, a rather strict form of interference control is being addressed under these limitations. Now, including the peak power constraint, we have
\begin{align}
P_i \le \min\left\{P_{\pk}, \frac{Q_{\pk}}{|g|^2}\right\} \label{eq:instantaneous-power-g}
\end{align}
for $i= 0,1$ (while keeping $P_1 = 0$ in the OSA scheme). Above, $Q_{\pk}$ denotes the peak received power limit at the primary receiver.

%

\subsection{Cognitive Channel Model}
As a result of channel sensing decisions and the true nature of primary user activity, we have four possible cases which are described below together with corresponding input-output relationships:
\begin{itemize}
\item \emph{Case (\Rmnum{1})}: A busy channel is sensed as busy, denoted by the joint event $(\mH_{1}, \hH_{1})$.
\begin{equation}
\hspace{-2.5cm}\text{(Correct detection)}\hspace{0.6cm}y = hs + n + w.
\label{eq:received_signal1}
\end{equation}
\item \emph{Case (\Rmnum{2})}: A busy channel is sensed as idle, denoted by the joint event $(\mH_{1}, \hH_{0})$.
\begin{equation}
\hspace{-2.4cm}\text{(Miss-detection)}\hspace{1cm}y = hs + n + w .
\label{eq:received_signal2}
\end{equation}
\item \emph{Case (\Rmnum{3})}: An idle channel is sensed as busy, denoted by the joint event $(\mH_{0}, \hH_{1})$.
\begin{equation}
\hspace{-3cm}\text{(False alarm)}\hspace{1.4cm}y = hs + n .
\label{eq:received_signal3}
\end{equation}
\item \emph{Case (\Rmnum{4})}: An idle channel is sensed as idle, denoted by the joint event $(\mH_{0}, \hH_{0})$.
\begin{equation}
\hspace{-3cm}\text{(Correct detection)}\hspace{0.6cm}y = hs + n .
\label{eq:received_signal4}
\end{equation}
\end{itemize}
In the above expressions, $s$ is the transmitted signal, $y$ is the received signal, and $h$ denotes zero-mean, circularly-symmetric complex fading coefficient between the secondary transmitter and the secondary receiver with variance $\sigma_h^2$. $n$ is the circularly-symmetric complex Gaussian noise with mean zero and variance $\E\{|n|^2\}= \sigma_{n}^2$, and hence has the pdf
\begin{align}
\label{eq:Gaussian_noise}
f_{n}(n) = \frac{1}{2\pi\sigma_n^2}\rme^{-\frac{|n|^2}{2\sigma_n^2}} = \frac{1}{2\pi\sigma_n^2}\rme^{-\frac{n_r^2 + n_i^2}{2\sigma_n^2}}.
\end{align}
The active primary user's received faded signal at the secondary receiver is denoted by $w$.  Notice that if the primary users are active and hence the hypothesis $\mH_1$ is true as in cases (\Rmnum{1}) and (\Rmnum{2}), the secondary receiver experiences interference from the primary user's transmission in the form of $w$. We assume that $w$ has a Gaussian mixture distribution, i.e., its pdf is a weighted sum of $p$ complex Gaussian distributions with zero mean and variance $\sigma_l^2$ for $1 \le l \le p$:
\begin{align}
\label{eq:Gaussian_mix}
f_w(w) = \sum_{l=1}^p \frac{\lambda_l}{2\pi\sigma_l^2}\rme^{-\frac{|w|^2}{2\sigma_l^2}}
\end{align}
where the weights $\lambda_l$ satisfy $\sum_{l=1}^p \lambda_l = 1$ with $\lambda_l \ge 0$ for all $l$.

\begin{Rem}
Primary user's received faded signal has a Gaussian mixture distribution, if we, for instance, have
\begin{align}
w=h_{ps}u
\end{align}
where $h_{ps}$, which is the channel fading coefficient between the primary transmitter and the secondary receiver, is a circularly symmetric, complex, zero-mean, Gaussian random variable, and $u$ is the primary user's modulated digital signal. Note that $w$ is conditionally Gaussian given $u$. Now, assuming that the modulated signal $u$ can take $p$ different values with prior probabilities given by $\lambda_l$ for $1 \le l \le p$, $w$ has a Gaussian mixture distribution as in (\ref{eq:Gaussian_mix}). In the case of multiple primary transmitters for which we have
\begin{align}
w=\sum_{i} h_{ps,i}u_i,
\end{align}
the above argument can easily be extended if all channel fading coefficients $\{h_{ps,i}\}$ are zero-mean Gaussian distributed.
\end{Rem}

\begin{Rem}
Gaussian mixture model is generally rich enough to accurately approximate a wide variety of density functions \cite[Section 3.2]{Stergiopoulos}. This fact indicates that the applicability of our results can be extended to various other settings in which $w$ has a distribution included in this class of densities.
Additionally, in the special case of $p = 1$, the Gaussian mixture distribution becomes the pure complex Gaussian distribution. Hence, the results obtained for the Gaussian mixture distribution can readily be specialized to derive those for the Gaussian distributed $w$ as well.
\end{Rem}

As observed from the input-output relationships in (\ref{eq:received_signal1})--(\ref{eq:received_signal4}), when the true state of the primary users is idle, corresponding to the cases (\Rmnum{3}) and (\Rmnum{4}), the additive disturbance is simply the background noise $n$. On the other hand, in cases (\Rmnum{1}) and (\Rmnum{2}) in which the channel is actually busy, the additive disturbance becomes
\begin{align} \label{eq:noise+int}
z = n + w \hspace{1 cm} \text{if } \mH_1 \text{ is true}
\end{align}
whose distribution can be obtained through the convolution of density functions of the background Gaussian noise $n$ and the primary user's received faded signal $w$. Using the result of Gaussian convolution of Gaussian mixture given by \cite{Pulkkinen}, the distribution of $z$ can be obtained as
\begin{align}
\label{eq:Gaussian_mix_conv}
f_z(z) = \sum_{l=1}^p \frac{\lambda_l}{2\pi(\sigma_l^2 + \sigma_n^2)}\rme^{-\frac{|z|^2}{2(\sigma_l^2 + \sigma_n^2)}}.
\end{align}
Note that $z$ also has a Gaussian mixture distribution. Note further that the pdf of $z$ can be expressed in terms of its real and imaginary components as
\begin{align}
\label{eq:Gaussian_mix_conv}
f_{z_r, z_i}(z_r, z_i) = \sum_{l=1}^p \frac{\lambda_l}{2\pi(\sigma_l^2 + \sigma_n^2)}\rme^{-\frac{(z_r + z_i)^2}{2(\sigma_l^2 + \sigma_n^2)}}.
\end{align}
Moreover, the marginal distributions of each component are given by
\begin{align}
\label{eq:marginal_inphase}
f_{z_r}(z_r) = \sum_{l=1}^{p} \frac{\lambda_l}{\sqrt{2\pi(\sigma_l^2 + \sigma_n^2)}}\rme^{-\frac{z_r^2}{2(\sigma_l^2 + \sigma_n^2)}}, \\
\label{eq:marginal_quad}
f_{z_i}(z_i) = \sum_{l=1}^{p} \frac{\lambda_l}{\sqrt{2\pi(\sigma_l^2 + \sigma_n^2)}}\rme^{-\frac{z_i^2}{2(\sigma_l^2 + \sigma_n^2)}}.
\end{align}
It is easily seen that the pdf of $z$ in (\ref{eq:Gaussian_mix_conv}) cannot be factorized into the product of the marginal pdf's of its real and imaginary parts $f_{z_r}(z_r)f_{z_i}(z_i)$, given in (\ref{eq:marginal_inphase}) and (\ref{eq:marginal_quad}), respectively. Therefore, the real and imaginary parts of the additive disturbance $z$ are dependent. When $p = 1$, i.e., in the case of a pure Gaussian distribution, the joint distribution can written as a product of its real and imaginary components since they are independent.

\section{General Formulations for the Optimal Decision Rule and Error Probabilities}\label{sec:general_formulation}
In this section, we present the optimal decision rule and the average symbol error probability for the cognitive radio system in the presence of channel sensing errors. We provide general formulations applicable to any modulation type under SSS and OSA schemes. More specific analysis for arbitrary rectangular QAM and PAM is conducted in Section \ref{sec:error_rate_QAM_PAM}.

\subsection{The Optimal Decision Rule} \label{subsec:optimaldecision}

In the cognitive radio setting considered in this paper, the optimal maximum \textit{a posteriori} probability (MAP) decision rule given the sensing decision $\hH_k$ and the channel fading coefficient $h$ can be formulated for any arbitrary $M$-ary digital modulation as follows:
\begin{align}
\label{eq:MAP_decision_rule}
\hat{s} &= \argmax_{0\le m\le M-1}\, \Pr\{s_m|y,h,\hH_k\}
\\
&= \argmax_{0\le m\le M-1}\, p_mf(y | s_m, h, \hH_k) \label{eq:MAP_decision_rule-2}
\\
&= \argmax_{0\le m\le M-1}\, p_m\big(\Pr\{\mH_0|\hH_{k}\}f(y | s_m, h, \hH_{k}, \mH_0) \nonumber \\ &\hspace{2.7cm}+\Pr\{\mH_1|\hH_{k}\}f(y | s_m, h, \hH_{k}, \mH_1)\big), \label{eq:cond-prob-yr}
\end{align}
where $p_m$ is the prior probability of signal constellation point $s_m$ and $k$ $\in \{0,1\}$. Above, (\ref{eq:MAP_decision_rule-2}) follows from Bayes' rule and (\ref{eq:cond-prob-yr}) is obtained by conditioning the density function on the hypotheses $\mH_0$ and $\mH_1$. Note that $f(y|s_m, h, \hH_k, \mH_j)$ in (\ref{eq:cond-prob-yr}) is the conditional distribution of the received real signal $y$ given the transmitted signal $s_m$, channel fading coefficient $h$, channel sensing decision $\hH_k$, and true state of the channel $\mH_j$, and can be expressed as
\begin{align}
f(y|s_m, h, \hH_k, \mH_j) = \begin{cases} \frac{1}{2\pi\sigma_n^2}\rme^{-\frac{|y - s_mh|^2}{2\sigma_n^2}}, &j=0\\
\sum_{l=1}^p \frac{\lambda_l}{2\pi(\sigma_l^2 + \sigma_n^2)}\rme^{-\frac{|y - s_mh|^2}{2(\sigma_l^2 + \sigma_n^2)}}, &j=1
\end{cases}
\label{eq:distrubition_yr}
\end{align}
for $m = 0,\dots,M-1$. Note that the sensing decision $\hH_k$ affects the density function through the power of the transmitted signal $s_m$.  Moreover, the conditional probabilities in (\ref{eq:cond-prob-yr}) can be expressed as
\begin{align}
\Pr\{\mathcal{H}_j | \hat{\mathcal{H}}_k\} = \frac{\Pr\{\mathcal{H}_j\} \Pr\{\hat{\mathcal{H}}_k | \mathcal{H}_j\}}{\sum_{j = 0}^1 \Pr\{\mathcal{H}_j\} \Pr\{\hat{\mathcal{H}}_k | \mathcal{H}_j\}} \quad j, k \in \{0,1\}, \nonumber
\end{align}
where $\Pr\{\mH_0\}$ and $\Pr\{\mH_1\}$ are the prior probabilities of the channel being idle and busy, respectively, and the conditional probabilities in the form $\Pr\{\hat{\mathcal{H}}_k | \mathcal{H}_j\}$ depend on the channel sensing performance. As discussed in Section \ref{subsec:sensing}, $P_{\nid} = \Pr\{\hH_1 | \mH_1\}$ is the detection probability and $P_{\f} = \Pr\{\hH_1 | \mH_0\}$ is the false alarm probability. From these formulations, we see that the optimal decision rule in general depends on the sensing reliability.


\subsection{Average Symbol Error Probability} \label{subsec:errorprob}

The average symbol error probability (SEP) for the MAP decision rule in (\ref{eq:MAP_decision_rule}) in the SSS scheme can be computed as
\begin{equation}
\hspace{-0.8cm}\begin{split}
&\SEP = 1 - \sum_{m=0}^{M-1}p_m\Pr\{\hat{s} = s_m|s_m\} \\
&= 1 - \sum_{m=0}^{M-1}\sum_{k=0}^{1}p_m\Pr\{\hH_k\}\Pr\{\hat{s} = s_m|s_m,\hH_k\}\\
&\label{eq:Pe_general}= 1 - \sum_{m=0}^{M-1}\sum_{j,k=0}^{1}p_m\Pr\{\hH_k\}\Pr\{\mH_j|\hH_k\}\Pr\{\hat{s} = s_m|s_m,\hH_k,\mH_j\}.
\end{split}
\end{equation}
The above average symbol error probability can further be expressed as in (\ref{eq:Pe_intdecisionregions})
\begin{figure*}
\begin{align}
\small
\begin{split}
\label{eq:Pe_intdecisionregions}
\SEP = 1 - \sum_{m=0}^{M-1}p_m\Bigg[\Pr\{\hH_0\}\bigg(\Pr\{\mH_1|\hH_0\}\int_{D_{m,0}} f(y|s_m, h, \hH_0, \mH_1)\,dy+ \Pr\{\mH_0|\hH_0\}\int_{D_{m,0}} f(y|s_m, h, \hH_0, \mH_0)\,dy\bigg)
\\+\Pr\{\hH_1\}\bigg(\Pr\{\mH_1|\hH_1\}\int_{D_{m,1}} f(y|s_m, h, \hH_1, \mH_1)\,dy
+ \Pr\{\mH_0|\hH_1\}\int_{D_{m,1}} f(y|s_m, h, \hH_1, \mH_0)\,dy\bigg)\Bigg]
\end{split}
\end{align}
\normalsize
\hrule
\end{figure*}
where $D_{m,0}$ and $D_{m,1}$ are the decision regions of each signal constellation point $s_m$ for $0\le m\le M-1$ when the channel is sensed to be idle and busy, respectively.

If cognitive user transmission is not allowed in the case of the channel being sensed as occupied by the primary users, we have the OSA scheme for which the average probability of error can be expressed as
\begin{equation}
\begin{split}
&\SEP =1 \!-\! \sum_{m=0}^{M-1}\sum_{j=0}^{1}p_m\bigg(\!\Pr\{\mH_j|\hH_0\}\Pr\{\hat{s} = s_m|s_m,\hH_0,\mH_j\}\!\bigg)\\
&= 1 - \sum_{m=0}^{M-1}\sum_{j=0}^{1}p_m\bigg(\Pr\{\mH_j|\hH_0\}\int_{D_{m,0}} f(y|s_m, h,\hH_0,\mH_j)\bigg).
\end{split}
\end{equation}

\section{Error Rate Analysis for $M$-ary Rectangular QAM}\label{sec:error_rate_QAM_PAM}

In this section, we conduct a more detailed analysis by considering rectangular QAM transmissions to demonstrate the key tradeoffs in a lucid setting. Correspondingly, we determine the optimal decision regions by taking channel sensing errors into consideration and identify the error rates for SSS and OSA schemes. We derive closed-form minimum average symbol error probability expressions under the transmit power and interference constraints. Note that the results for QAM can readily be specialized for PAM, QPSK, and BPSK transmissions.

\subsection{Optimal decision regions under channel sensing uncertainty}
The signal constellation point $s_{i,q}$ in $M_I\times M_Q$ rectangular QAM signaling can be expressed in terms of its real and imaginary parts, respectively, as
\begin{align}
s_{n,q} = s_{n} + js_{q}, \label{eq:complexQAMpoint}
\end{align}
where the amplitude level of each component is given by
\begin{align}
s_{n} &= (2n+1-M_I)\frac{d_{min,k}}{2} \hspace{0.4 cm}\text{for } \hspace{0.2 cm} n = 0,\dots,M_I-1,  \label{eq:realQAMpoint}\\
s_{q} &= (2q+1-M_Q)\frac{d_{min,k}}{2} \hspace{0.4 cm} \text{for } \hspace{0.2 cm} q = 0,\dots,M_Q-1. \label{eq:imaginaryQAMpoint}
\end{align}
Above, $M_I$ and $M_Q$ are the modulation size on the in-phase and quadrature components, respectively, and $d_{min,k}$ denotes the minimum distance between the signal constellation points and is given by
\begin{align}
d_{min,k} = \sqrt{\frac{12P_k}{M_I^2 + M_Q^2-2}} \hspace{1 cm}k \in \{0, 1\}
\end{align}
where $P_k$ is the transmission power under sensing decision $\hH_{k}$.

It is assumed that the fading realizations are perfectly known at the receiver. In this case, phase rotations caused by the fading can be offset by multiplying the channel output $y$ with $\rme^{-j \theta_h}$ where $\theta_h$ is the phase of the fading coefficient $h$. Hence, the modified received signal can be written in terms of its real and imaginary parts as follows:
\begin{equation}
\begin{split}
\by &= \byr + j\byi = y\rme^{-j\theta_h} \\ &= \begin{cases} s_{n}|h| + \bn_r + j(s_{q}|h|+ \bni), &\text{ under } \mH_{0}\\
s_{n}|h| + \bnr + \bwr + j(s_{q}|h|+ \bni + \bwi), &\text{ under } \mH_{1}
\end{cases} \nonumber
\end{split}
\end{equation}
where the subscripts $r$ and $i$ are used to denote the real and imaginary components of the signal, respectively. Note that $\bn= \bnr + j\bni$ and $\bw = \bwr + j\bwi$ have the same statistics as $n$ and $w$, respectively, due to their property of being circularly symmetric.
Hence, given the transmitted signal constellation point $s_{n,q}$, the distribution of the modified received signal $\by$ is given by
\begin{equation}
\begin{split}
f_{\by_r, \by_i}&(\by_r, \by_i|s_{n,q}, h, \hH_k, \mH_j)  \\ &=\begin{cases} \frac{1}{2\pi\sigma_n^2}\rme^{-\frac{(\by_r - s_{n}|h|)^2+(\by_i - s_{q}|h|)^2}{2\sigma_n^2}}, &j=0\\
\sum_{l=1}^p \frac{\lambda_l}{2\pi(\sigma_l^2 + \sigma_n^2)}\rme^{-\frac{(\by_r - s_{n}|h|)^2+(\by_i - s_{q}|h|)^2}{2(\sigma_l^2 + \sigma_n^2)}}, &j=1
\end{cases}.
\label{eq:distrubition_modified_y}
\end{split}
\end{equation}
Moreover, the real and imaginary parts of noise, i.e., $\bnr$ and $\bni$ are independent zero-mean Gaussian random variables, and the real and imaginary parts of primary users' faded signal, i.e., $\bwr$ and $\bwi$, are Gaussian mixture distributed random variables.

In the following, we characterize the decision regions of the optimal detection rule for equiprobable QAM signaling in the presence of sensing uncertainty.

\begin{Prop} \label{prop:decisionrule}
For cognitive radio transmissions with \emph{equiprobable} rectangular $M$-QAM modulation (with constellation points expressed as in (\ref{eq:complexQAMpoint})--(\ref{eq:imaginaryQAMpoint})) under channel sensing uncertainty in both SSS and OSA schemes, the optimal detection thresholds under any channel sensing decision are located midway between the received signal points. Hence, the optimal detector structure does not depend on the sensing decision.
\end{Prop}

\emph{Proof}: See Appendix \ref{app:proof-Prop}.

\subsection{The average symbol error probability under channel sensing uncertainty}
In this subsection, we present closed-form average symbol error probability expressions under both transmit power and interference constraints for SSS and OSA schemes. Initially, we express the error probabilities for a given value of the fading coefficient $h$. Subsequently, we address averaging over fading and also incorporate power and interference constraints. We note that in the presence of peak interference constraints, the transmitted power level depends on the fading coefficient $g$ experienced in the channel between the secondary and primary users as seen in (\ref{eq:instantaneous-power-g}). Therefore, we in this case consider an additional averaging of the error rates with respect to $g$.

\subsubsection{Sensing-based spectrum sharing (SSS) scheme} \label{subsubsec:error_rate_SSS}
Under the optimal decision rule given in the previous subsection, the average symbol error probability of \emph{equiprobable} signals for a given fading coefficient $h$ can be expressed as
\begin{equation}
\begin{split}
&\SEP(P,h)
\\
&= \sum_{m=1}^{M} \sum_{j,k=0}^{1}\frac{\Pr\{\hH_k\}}{M}\Pr\{\mH_j|\hH_k\}\Pr\{e|s_{n,q},h,\hH_k,\mH_j\},
\label{eq:Pe_MPAM_h}
\end{split}
\end{equation}
where $\Pr\{e|s_{n,q},h,\mH_j,\hH_k \}$ denotes the conditional error probability given the transmitted signal $s_{n,q}$, channel fading $h$, sensing decision $\hH_k$, and true state of the channel $\mH_j$.

We can group the error patterns of rectangular $M$-QAM modulation into three categories. Specifically, the probability of error for the signal constellation points on the corners is equal due to the symmetry in signaling and detection. The same is also true for the points on the sides and the inner points.

The symbol error probability for the four corner points is given by
\begin{equation}
\begin{split}
\SEP_{1,k}(P,h) =1 - \int_{a_1}^\infty \int_{a_1}^\infty &\big(\!\Pr\{\!\mH_0|\hH_k\} f_{n_r,n_i}(n_r, \!n_i)dn_rdn_i \\ +&\Pr\{\!\mH_1|\hH_k\} f_{z_r,z_i}(z_r,z_i)dz_rdz_i\big)
\end{split}
\end{equation}
where $a_1=-\frac{d_{min,k}|h|}{2}$ and $k \in \{0,1\}$. The distributions of the Gaussian noise $f_{n_r,n_i}(n_r, n_i) $ and the primary user's interference signal plus noise $f_{z_r,z_i}(z_r, z_i)$ are given in (\ref{eq:Gaussian_noise}) and (\ref{eq:Gaussian_mix_conv}), respectively. After evaluating the integrals, the above expression becomes
\begin{equation}
\small
\begin{split}\label{eq:SEP1}
&\SEP_{1,k}(P,h)
\\
&= \Pr\{\mH_0|\hH_k\}\Bigg\{\!2Q\bigg(\sqrt{\frac{d_{min,k}^2|h|^2}{4\sigma_n^2}}\bigg) - Q^2\bigg(\sqrt{\frac{d_{min,k}^2|h|^2}{4\sigma_n^2}}\bigg)\!\Bigg\} \\&+\Pr\{\mH_1|\hH_k\}\sum_{l=1}^p\lambda_l \Bigg\{\!2Q\bigg(\sqrt{\frac{d_{min,k}^2|h|^2}{4(\sigma_l^2+\sigma_n^2)}}\bigg) - Q^2\bigg(\sqrt{\frac{d_{min,k}^2|h|^2}{4(\sigma_l^2+\sigma_n^2)}}\bigg)\!\Bigg\}
\end{split}
\end{equation}
\normalsize
where $Q(x) = \int_{x}^{\infty}\frac{1}{\sqrt{2\pi}}e^{-t^2/2}dt$ is the Gaussian $Q$-function. For the $2(M_I + M_Q -4)$ points on the sides, except the corner points, the symbol error probability is
\begin{equation}
\begin{split}
\SEP_{2,k}(P,h)=1 - \int_{a_1}^\infty \int_{a_1}^{a_2}&\big(\Pr\{\mH_0|\hH_k\} f_{n_r,n_i}(n_r, n_i)dn_rdn_i \\+ &\Pr\{\mH_1|\hH_k\!\} f_{z_r,z_i}(z_r, z_i)dz_rdz_i\big)
\end{split}
\end{equation}
where $a_2=\frac{d_{min,k}|h|}{2}$. After performing the integrations, we can express $\SEP_{2,k}(P,h)$ as
\begin{equation}
\small
\begin{split}\label{eq:SEP2}
&\SEP_{2,k}(P,h)
\\&=
\Pr\{\mH_0|\hH_k\}\Bigg\{\!3Q\bigg(\sqrt{\frac{d_{min,k}^2|h|^2}{4\sigma_n^2}}\bigg) - 2Q^2\bigg(\sqrt{\frac{d_{min,k}^2|h|^2}{4\sigma_n^2}}\bigg)\!\Bigg\} \\&\hspace{-0.2cm}+\Pr\{\mH_1|\hH_k\}\sum_{l=1}^p\lambda_l \Bigg\{\!3Q\bigg(\sqrt{\frac{d_{min,k}^2|h|^2}{4(\sigma_l^2+\sigma_n^2)}}\bigg) - 2Q^2\bigg(\sqrt{\frac{d_{min,k}^2|h|^2}{4(\sigma_l^2+\sigma_n^2)}}\bigg)\!\Bigg\}.
\end{split}
\end{equation}
\normalsize
Finally, the symbol error probability for $M- 2(M_I+M_Q)+4$ inner points is obtained from
\begin{equation}
\begin{split}
\SEP_{3,k}(P,h)=1 - \int_{a_1}^{a_2} \int_{a_1}^{a_2} &\big(\Pr\{\mH_0|\hH_k\} f_{n_r,n_i}(n_r, n_i)dn_rdn_i \\+ &\Pr\{\mH_1|\hH_k\} f_{z_r,z_i}(z_r, z_i)dz_rdz_i\big)
\end{split}
\end{equation}
which can be evaluated to obtain
\begin{equation}
\small
\begin{split}\label{eq:SEP3}
&\SEP_{3,k}(P,h)
\\
&= \Pr\{\mH_0|\hH_k\}\Bigg\{\!4Q\bigg(\sqrt{\frac{d_{min,k}^2|h|^2}{4\sigma_n^2}}\bigg) - 4Q^2\bigg(\sqrt{\frac{d_{min,k}^2|h|^2}{4\sigma_n^2}}\bigg)\!\Bigg\} \\&\hspace{0.1cm}+\Pr\{\mH_1|\hH_k\}\sum_{l=1}^p\lambda_l \Bigg\{\!4Q\bigg(\sqrt{\frac{d_{min,k}^2|h|^2}{4(\sigma_l^2+\sigma_n^2)}}\bigg) - 4Q^2\bigg(\sqrt{\frac{d_{min,k}^2|h|^2}{4(\sigma_l^2+\sigma_n^2)}}\bigg)\!\Bigg\}.
\end{split}
\end{equation}
\normalsize
Overall, we can express $\SEP(P,h)$ in (\ref{eq:Pe_MPAM_h}) by combining $\SEP_{1,k}(P,h)$, $\SEP_{2,k}(P,h)$ and $\SEP_{3,k}(P,h)$ as follows
\begin{equation}
\begin{split}
\SEP(P,h) = &\sum_{k=0}^{1}\frac{\Pr\{\hH_k\}}{M}\Big(4 \, \SEP_{1,k}(P,h)\\&+2(M_I + M_Q -4)\SEP_{2,k}(P,h) \\ &+ (M- 2(M_I+M_Q)+4)\SEP_{3,k}(P,h) \Big).
\end{split}
\end{equation}
\normalsize
After rearranging the terms, the final expression for the average symbol error probability $\SEP(P,h)$ is given by (\ref{eq:Pe_h_QAM}) shown at the top of next page.
\begin{figure*}
\begin{equation}
\small
\begin{split}
\label{eq:Pe_h_QAM}
\hspace{-0.8 cm}\SEP(P,h)=\sum_{k=0}^{1}\Pr\{\hH_k\} \Bigg\{\Pr\{\mH_0|\hH_k\}\Bigg[2\bigg(2-\frac{1}{M_I}-\frac{1}{M_Q}\bigg)Q\Bigg(\sqrt{\frac{d_{min,k}^2|h|^2}{4\sigma_{n}^2}} \Bigg)-4\bigg(1-\frac{1}{M_I}\bigg)\bigg(1-\frac{1}{M_Q}\bigg)Q^2\Bigg(\sqrt{\frac{d_{min,k}^2|h|^2}{4\sigma_{n}^2}} \Bigg)\Bigg]\\
+\Pr\{\mH_1|\hH_k\}\Bigg[2\bigg(2-\frac{1}{M_I}-\frac{1}{M_Q}\bigg)\sum_{l=1}^p\lambda_l Q\Bigg(\sqrt{\frac{d_{min,k}^2|h|^2}{4(\sigma_{l}^2+\sigma_{n}^2)}} \Bigg)-4\bigg(1-\frac{1}{M_I}\bigg)\bigg(1-\frac{1}{M_Q}\bigg)\sum_{l=1}^p\lambda_l Q^2\Bigg(\sqrt{\frac{d_{min,k}^2|h|^2}{4(\sigma_{l}^2+\sigma_{n}^2)}} \Bigg)\Bigg]\Bigg\}.
\end{split}
\end{equation}
\hrule
\end{figure*}
\normalsize
This expression can be specialized to square QAM signaling by setting $M_I = M_Q = \sqrt{M}$.

We observe above that while the optimal decision rule does not depend on the sensing decisions, the error rates are functions of detection and false alarm probabilities. Note also that the expressions above are for a given value of fading. The unconditional symbol error probability averaged over fading can be evaluated from
\begin{align}
\SEP(P) = \int_0^{\infty} \SEP(P,x)f_{|h|^2} (x)dx.
\end{align}
In the special case of a Rayleigh fading model for which the fading power has an exponential distribution with unit mean, i.e., $f_{|h|^2} (x) = \rme^{-x}$,  the above integral can be evaluated by adopting the same approach as in \cite{Alouini} and using the indefinite integral form of the Gaussian $Q$-function \cite{Craig} and square of the Gaussian $Q$-function \cite{Simon}, given, respectively, by
\begin{align}
\label{eq:Q_exp}
Q(x) &= \frac{1}{\pi}\int_0^{\frac{\pi}{2}}\rme^{\bigg(-\frac{x^2}{2sin^2\phi}\bigg)}d\phi,\\
\label{eq:Q2_exp} Q^2(x) &= \frac{1}{\pi}\int_0^{\frac{\pi}{4}}\rme^{\bigg(-\frac{x^2}{2sin^2\phi}\bigg)}d\phi  \hspace{0.3cm}\text{for }\hspace{0.3cm} x \geq 0.
\end{align}
The resulting unconditional average symbol error probability over Rayleigh fading is given by (\ref{eq:Pe_SSS_without_power_constraints}) at the top of next page
\begin{figure*}
\begin{equation}
\small
\begin{split}
\label{eq:Pe_SSS_without_power_constraints}
\hspace{-1 cm}\SEP(P) &=\sum_{k=0}^{1}\Pr\{\hH_k\}\Bigg\{\Pr\{\mH_0|\hH_k\}\Bigg[\bigg(2-\frac{1}{M_I}-\frac{1}{M_Q}\bigg)\bigg(1-\frac{1}{\beta_{0,k}}\bigg)-2\bigg(1-\frac{1}{M_I}\bigg)\bigg(1-\frac{1}{M_Q}\bigg)\bigg(\frac{2}{\pi}\frac{1}{\beta_{0,k}}\tan^{-1}\left(\frac{1}{\beta_{0,k}}\right)-\frac{1}{\beta_{0,k}}+\frac{1}{2}\bigg)\Bigg]\\
&+\Pr\{\mH_1|\hH_k\}\Bigg[\bigg(2-\frac{1}{M_I}-\frac{1}{M_Q}\bigg)\sum_{l=1}^p\lambda_l\bigg(1-\frac{1}{\beta_{1,k}}\bigg)-2\bigg(1-\frac{1}{M_I}\bigg)\bigg(1-\frac{1}{M_Q}\bigg)\sum_{l=1}^p\lambda_l\bigg(\frac{2}{\pi}\frac{1}{\beta_{1,k}}\tan^{-1}\left(\frac{1}{\beta_{1,k}}\right)-\frac{1}{\beta_{1,k}}+\frac{1}{2}\bigg)\Bigg]\Bigg\}.
\end{split}
\end{equation}
\hrule
\end{figure*}
\normalsize
where  $\beta_{0,k}=\sqrt{1+\frac{2}{3P_k}{(M_I}^2+{M_Q}^2-2)\sigma_{n}^2}$ and $\beta_{1,k}=\sqrt{1+\frac{2}{3P_k}({M_I}^2+{M_Q}^2-2)(\sigma_{l}^2 + \sigma_{n}^2)}$ for $1 \le l \le p$. The average symbol error probability for rectangular QAM signaling in the presence of Gaussian-distributed $w$ can readily be obtained by letting $l =1$ in (\ref{eq:Pe_SSS_without_power_constraints}). Although the SEP expression in (\ref{eq:Pe_SSS_without_power_constraints}) seems complicated, it is in fact very simple to evaluate. Furthermore, this $\SEP(P)$ can be upper bounded as
\begin{equation}
\small
\begin{split}
\label{eq:Pe_SSS_tight}
\SEP(P)\!\le\! \bigg(2\!-\frac{1}{M_I}\!-\frac{1}{M_Q}\bigg)\sum_{k=0}^{1}&\Pr\{\hH_k\}\Bigg\{\!\Pr\{\mH_0|\hH_k\}\bigg(1-\frac{1}{\beta_{0,k}}\bigg) \\ &\hspace{-0.2cm}+\Pr\{\mH_1|\hH_k\}\sum_{l=1}^p\lambda_l\bigg(1-\frac{1}{\beta_{1,k}}\bigg)\Bigg\}.
\end{split}
\end{equation}
\normalsize
This upper bound follows by removing the negative terms that include $Q^2(\cdot)$ on the right-hand side of (\ref{eq:Pe_h_QAM}) and then integrating with respect to fading distribution. Note also that the upper bound in (\ref{eq:Pe_SSS_tight}) with $M_Q=1$ is the exact symbol error probability for PAM modulation.

Note further that the SEP expression in (\ref{eq:Pe_SSS_without_power_constraints}) is a function of the transmission powers $P_0$ and $P_1$. The optimal choice of the power levels under peak power and average interference constraints given in (\ref{eq:peak_power_constraint}) and (\ref{eq:interference_power_constraint_SSS}) and the resulting error rates can be determined by solving
\begin{equation}
\small
\begin{split}
\label{eq:SSS_Expectation_Pe}
\SEP(P_{\pk}, Q_{\avg})= \min_{
\substack{P_0 \le P_{\pk}, \, P_1 \le P_{\pk} \\ (1-P_{\nid})\,P_0 \,\E\{|g|^2\} + P_{\nid} \,P_1 \, \E\{|g|^2\} \le Q_{\avg}}} \SEP(P_0, P_1).
\end{split}
\normalsize
\end{equation}

As discussed in Section \ref{subsec:power-interference}, if the fading coefficient $g$ between the secondary transmitter and the primary receiver is known and peak interference constraints are imposed, then the maximum transmission power is given by
\begin{align}
\label{eq:optimal_power_SSS}
P_i^*=\min\left\{P_{\pk},\frac{Q_{\pk}}{|g|^2}\right\} \quad \text{for } i=0,1.
\end{align}
After inserting these $P_0^*$  and $P_1^*$ into the $\SEP$ upper bound in (\ref{eq:Pe_SSS_tight}) and evaluating the expectation over the fading coefficient $g$, we obtain
\begin{equation}
\small
\begin{split}
\label{eq:Pe_SSS_tight-avg}
\SEP \le \int_0^{b_1}\SEP_u(P_{peak})f_{|g|^2}(y)dy+\int_{b_1}^{\infty}\SEP_u\bigg(\frac{Q_{\pk}}{y}\bigg)f_{|g|^2}(y)dy
\end{split}
\normalsize
\end{equation}
where $b_1=\frac{Q_{\pk}}{P_{\pk}}$ and $\SEP_u$ denotes the upper bound in (\ref{eq:Pe_SSS_tight}). If $|g|^2$ is exponentially distributed with unit mean, then by using the identity in \cite[eq. 3.362.2]{gradshteyn}, we can evaluate the second integral on the right-hand side of (\ref{eq:Pe_SSS_tight-avg}) and express the upper bound as in (\ref{eq:Pe_SSS_min}) given on the next page
\begin{figure*}
\begin{equation}
\small
\begin{split}
\label{eq:Pe_SSS_min}
\SEP \le (1-\rme^{-b_1})\SEP(P_{\pk})+\bigg(2-\frac{1}{M_I}-\frac{1}{M_Q}\bigg)\sum_{k=0}^{1}\Pr\{\hH_k\}\Bigg\{&\Pr\{\mH_0|\hH_k\}\bigg[\rme^{b_1}-2\sqrt{\gamma_{0}\pi}\rme^{\gamma_{0}}Q\big(\sqrt{2}(b_1+\gamma_{0})\big)\bigg]\\&+\Pr\{\mH_1|\hH_k\}\sum_{l=1}^p\lambda_l\bigg[\rme^{b_1}-2\sqrt{\gamma_{1}\pi}\rme^{\gamma_{1}}Q\big(\sqrt{2}(b_1+\gamma_{1})\big)\bigg]\Bigg\}\\
\end{split}
\end{equation}
\hrule
\end{figure*}
\normalsize
where $\gamma_{0}=\frac{3b_1P_{\pk}}{2(M_I^2+M_Q^2-2)\sigma_n^2}$, $\gamma_{1}=\frac{3b_1P_{\pk}}{2(M_I^2+M_Q^2-2)(\sigma_l^2+\sigma_n^2)}$.

It should be noted that we can easily obtain the \emph{exact} symbol error probability expressions for PAM modulation by replacing $M_I = M$ and $M_Q=1$ in (\ref{eq:Pe_h_QAM}), (\ref{eq:Pe_SSS_without_power_constraints}), (\ref{eq:Pe_SSS_min}).

\subsubsection{Opportunistic spectrum access (OSA) scheme}\label{subsubsec:error_rate_OSA}
In the OSA scheme, secondary users are not allowed to transmit when the primary user activity is sensed in the channel. Therefore, we only consider error patterns under $\hH_0$ given in (\ref{eq:SEP1}), (\ref{eq:SEP2}), (\ref{eq:SEP3}). Hence, following the same approach adopted in Section \ref{subsubsec:error_rate_SSS}, the average symbol error probability under the OSA scheme is obtained as in (\ref{eq:Pe_OSA_without_power_constraints}) given on the next page.
\begin{figure*}
\begin{equation}
\small
\begin{split}
\label{eq:Pe_OSA_without_power_constraints}
\hspace{-.5 cm}\SEP(P_0) &=\Pr\{\mH_0|\hH_0\}\Bigg[\bigg(2-\frac{1}{M_I}-\frac{1}{M_Q}\bigg)\bigg(1-\frac{1}{\beta_{0,0}}\bigg)-2\bigg(1-\frac{1}{M_I}\bigg)\bigg(1-\frac{1}{M_Q}\bigg)\bigg(\frac{2}{\pi}\frac{1}{\beta_{0,0}}\tan^{-1}\left(\frac{1}{\beta_{0,0}}\right)-\frac{1}{\beta_{0,0}}+\frac{1}{2}\bigg)\Bigg]\\
&+\Pr\{\mH_1|\hH_0\}\Bigg[\bigg(2-\frac{1}{M_I}-\frac{1}{M_Q}\bigg)\sum_{l=1}^p\lambda_l\bigg(1-\frac{1}{\beta_{1,0}}\bigg)-2\bigg(1-\frac{1}{M_I}\bigg)\bigg(1-\frac{1}{M_Q}\bigg)\sum_{l=1}^p\lambda_l\bigg(\frac{2}{\pi}\frac{1}{\beta_{1,0}}\tan^{-1}\left(\frac{1}{\beta_{1,0}}\right)-\frac{1}{\beta_{1,0}}+\frac{1}{2}\bigg)\Bigg].
\end{split}
\end{equation}
\hrule
\end{figure*}
Similarly, the SEP upper bound becomes
\begin{equation}
\small
\begin{split}
\label{eq:Pe_OSA_tight}
\SEP(P) \le \bigg(2-\frac{1}{M_I}-\frac{1}{M_Q}\bigg)&\bigg\{\Pr\{\mH_0|\hH_0\}\bigg(1-\frac{1}{\beta_{0,0}}\bigg) \\ &+\Pr\{\mH_1|\hH_0\}\sum_{l=1}^p\lambda_l\bigg(1-\frac{1}{\beta_{1,0}}\bigg)\bigg\}.
\end{split}
\end{equation}
\normalsize
Note that under average interference constraints, the maximum allowed transmission power in an idle-sensed channel is given by
\begin{gather}
P_0^* = \min\left\{ P_{\pk}, \frac{Q_{\avg}}{(1-P_{\nid}) \E\{|g|^2\}}\right\}. \label{eq:maxpower-averageinterference-OSA}
\end{gather}
On the other hand, if the peak interference power constraint is imposed, the maximum allowed transmission power is
\begin{align}
\label{eq:optimal_power_OSA}
P_0^*=\min\left\{P_{\pk},\frac{Q_{\pk}}{|g|^2}\right\}.
\end{align}
After inserting this $P_0^*$ into (\ref{eq:Pe_OSA_tight}), assuming again that $|g|^2$ is exponentially distributed with unit mean, and evaluating the integration in a similar fashion as in Section \ref{subsubsec:error_rate_SSS}, an upper bound on the average symbol error probability can be obtained as in (\ref{eq:Pe_OSA_min}) on the next page where $\psi_0=\frac{3Q_{pk}}{2(M_I^2+M_Q^2-2)\sigma_n^2}$ and $\psi_1=\frac{3Q_{pk}}{2(M_I^2+M_Q^2-2)(\sigma_l^2+\sigma_n^2)}$.
\begin{figure*}
\begin{equation}
\small
\begin{split}
\label{eq:Pe_OSA_min}
\SEP \le \left(1-\rme^{\frac{Q_{\pk}}{P_{\pk}}}\right)\SEP(P_{\pk})+\bigg(2-\frac{1}{M_I}-\frac{1}{M_Q}\bigg)\Bigg\{&\Pr\{\mH_0|\hH_0\}\bigg[\rme^{\frac{Q_{\pk}}{P_{\pk}}}-2\sqrt{\psi_{0}\pi}\rme^{\psi_{0}}Q\bigg(\sqrt{2}\bigg(\frac{Q_{\pk}}{P_{\pk}}+\psi_{0}\bigg)\bigg)\bigg]\\&+\Pr\{\mH_1|\hH_0\}\sum_{l=1}^p\lambda_l\bigg[\rme^{\frac{Q_{\pk}}{P_{\pk}}}-2\sqrt{\psi_{1}\pi}\rme^{\psi_{1}}Q\bigg(\sqrt{2}\bigg(\frac{Q_{\pk}}{P_{\pk}}+\psi_{1}\bigg)\bigg)\bigg]\Bigg\} \end{split}
\end{equation}
\hrule
\end{figure*}
\normalsize

\section{Numerical Results}\label{sec:num_results}
In this section, we present numerical results to demonstrate the error performance of a cognitive radio system in the presence of channel sensing uncertainty for both SSS and OSA schemes.  More specifically, we numerically investigate the impact of sensing performance (e.g., detection and false-alarm probabilities), different levels of peak transmission power and average and peak interference constraints on cognitive transmissions in terms of symbol error probability. Theoretical results are validated through Monte Carlo simulations. Unless mentioned explicitly, the following parameters are employed in the numerical computations. It is assumed that the variance of the background noise is $\sigma_{n}^2 = 0.01$. When the primary user signal is assumed to be Gaussian, its variance, $\sigma_{w}^2$ is set to 0.5. On the other hand, in the case of primary user's received signal $w$ distributed according to the Gaussian mixture model, we assume that $p = 4$, i.e., there are four components in the mixture, $\lambda_l = 0.25$ for all $1 \le l \le 4$, and the variance is still $\sigma_{w}^2 = 0.5$.
Also, the primary user is active over the channel with a probability of $0.4$, hence $\Pr\{\mH_1\} = 0.4$ and $\Pr\{\mH_0\} = 0.6$. Finally, we consider a Rayleigh fading channel between the secondary users with fading power pdf given by $f_{|h|^2} (x) = \rme^{-x}$ for $x \ge 0$, and also assume that the fading power $|g|^2$ in the channel between the secondary transmitter and primary receiver is exponentially distributed with $\E\{|g|^2\} = 1$.

\subsection{SEP under Average Interference Constraints} \label{subsec:averageinterference}

\begin{figure}
\centering
\begin{subfigure}[b]{0.45\textwidth}
\centering
\includegraphics[width=\textwidth]{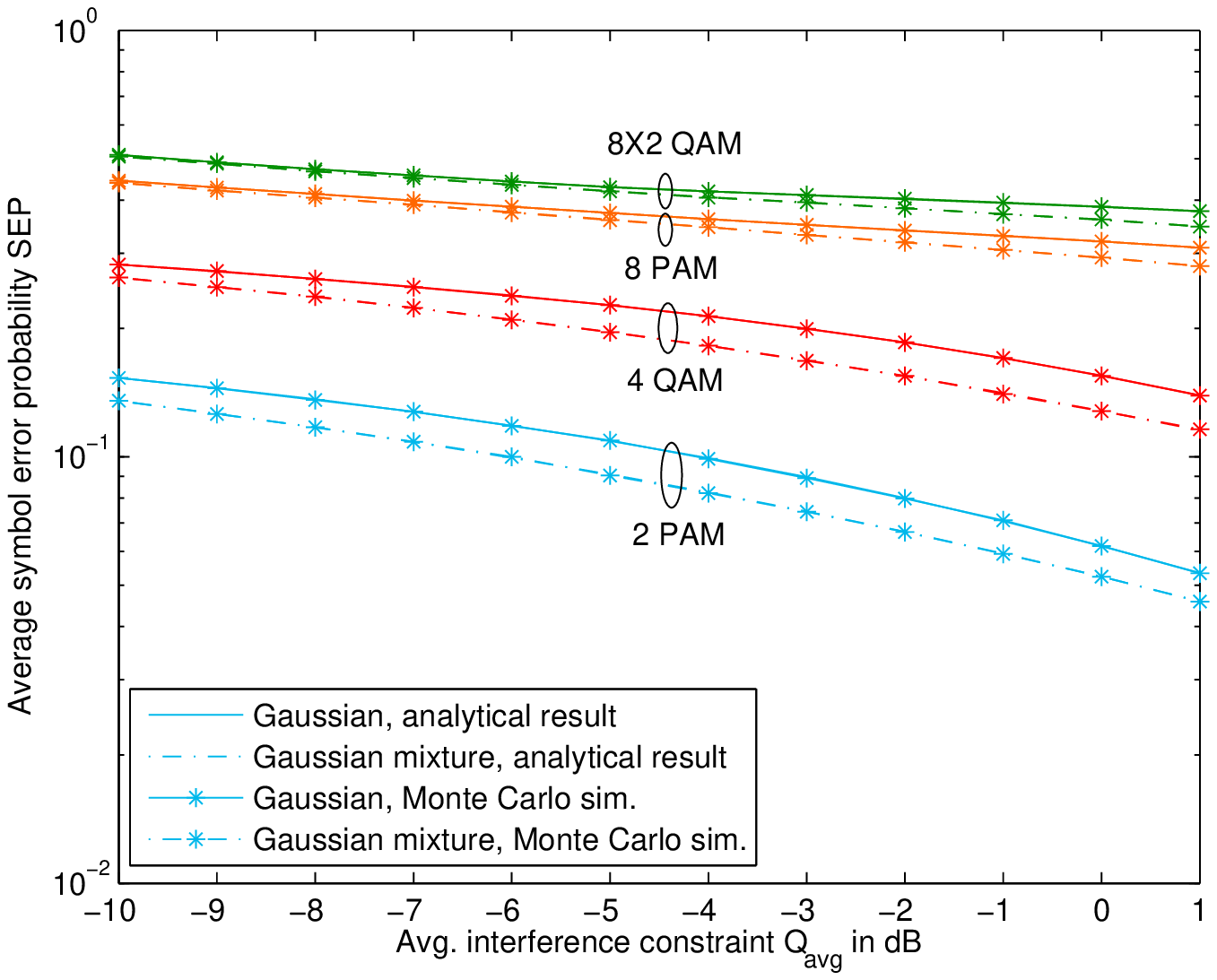}
\end{subfigure}
\begin{subfigure}[b]{0.45\textwidth}
\centering
\includegraphics[width=\textwidth]{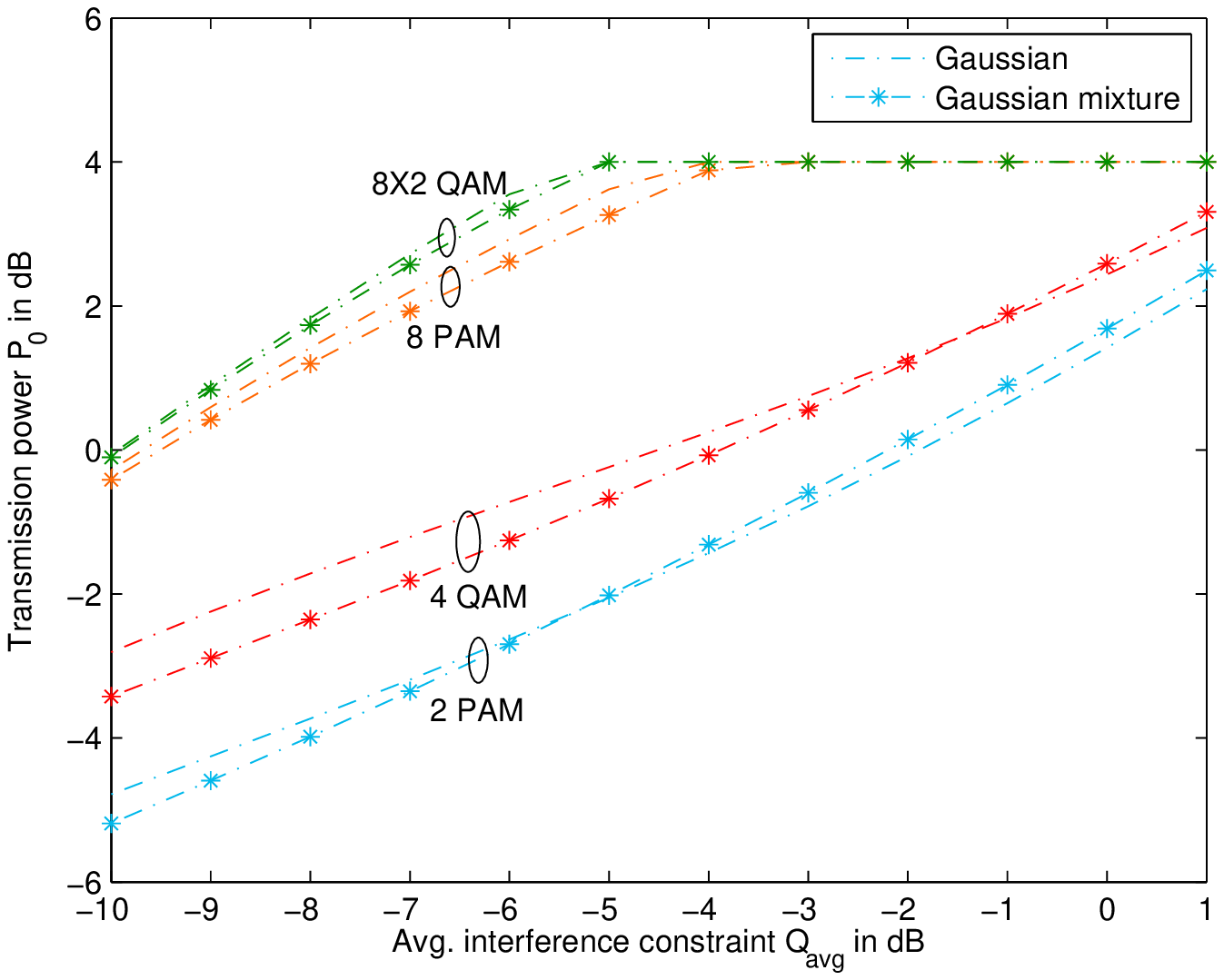}
\end{subfigure}
\begin{subfigure}[b]{0.45\textwidth}
\centering
\includegraphics[width=\textwidth]{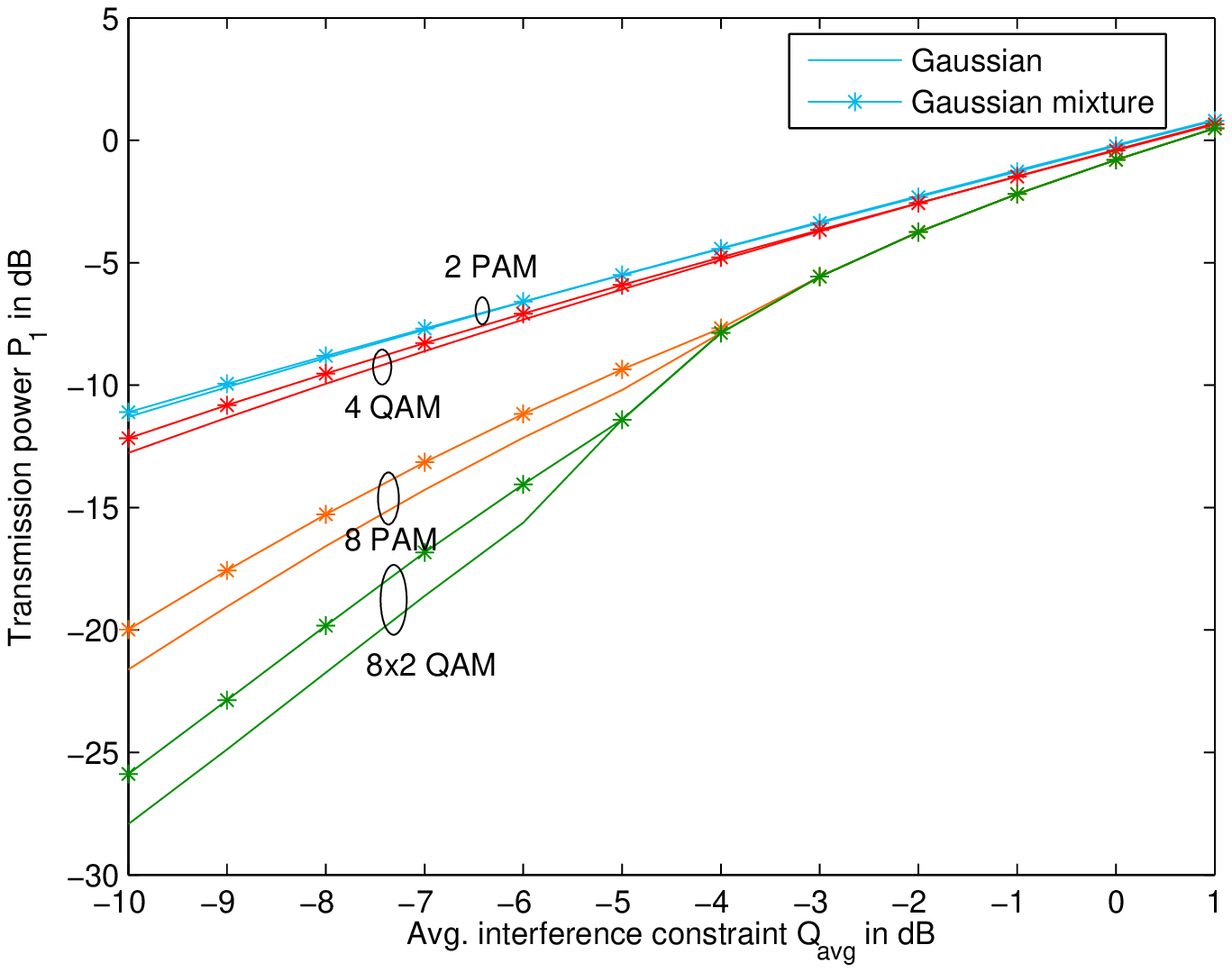}
\end{subfigure}
\caption{\small{Average symbol error probability SEP, and transmission powers $P_0$ and $P_1$ vs. average interference constraint, $Q_{\avg}$ in SSS scheme.}}\label{fig:SEP_P0_P1_Qavg_sss}
\end{figure}

We initially consider peak transmit and average interference constraints as given in (\ref{eq:peak_power_constraint}) and (\ref{eq:interference_power_constraint_SSS}), respectively. In the following numerical results, for the SSS scheme, we plot the error probabilities and optimal transmission power levels obtained by solving (\ref{eq:SSS_Expectation_Pe}). In  the case of OSA, we plot the average error probability expressed in (\ref{eq:Pe_OSA_without_power_constraints}) with maximum allowed power $P_0^*$ given in (\ref{eq:maxpower-averageinterference-OSA}).

In Fig. \ref{fig:SEP_P0_P1_Qavg_sss}, we display the average symbol error probability ($\SEP$) and optimal transmission powers $P_0$ and $P_1$ as a function of the average interference constraint, $Q_{\avg}$, in the SSS scheme.  $P_{\nid}$ and $P_{\f}$ are set to $0.9$ and $0.05$, respectively. The peak transmission power is $P_{\pk} = 4$ dB. We assume that the secondary users employ 2-PAM, 4-QAM, 8-PAM and $8\times 2$-QAM modulation schemes for data transmission. We have considered both Gaussian and Gaussian-mixture distributed $w$. In addition to the analytical results obtained by using (\ref{eq:Pe_SSS_without_power_constraints}) and solving (\ref{eq:SSS_Expectation_Pe}), we performed Monte Carlo simulations to determine the SEP. We notice in the figure that analytical and simulation results agree perfectly.
Additionally, it is seen that for all modulation types, error rate performance of secondary users improves as average interference constraint becomes looser (i.e., as $Q_{\avg}$ increases), allowing transmission power levels $P_0$ and $P_1$ to become higher as illustrated in the lower subfigures. Saturation seen in the plot of $P_0$ is due to the peak constraint $P_{\pk}$. Other observations are as follows. As the modulation size increases, $\SEP$ increases as expected. It is also interesting to note that lower $\SEP$ is attained in the presence of Gaussian-mixture distributed $w$ when compared with the performance when $w$ has a pure Gaussian density with the same variance.

\begin{figure}
\centering
\begin{subfigure}[b]{0.45\textwidth}
\centering
\includegraphics[width=\textwidth]{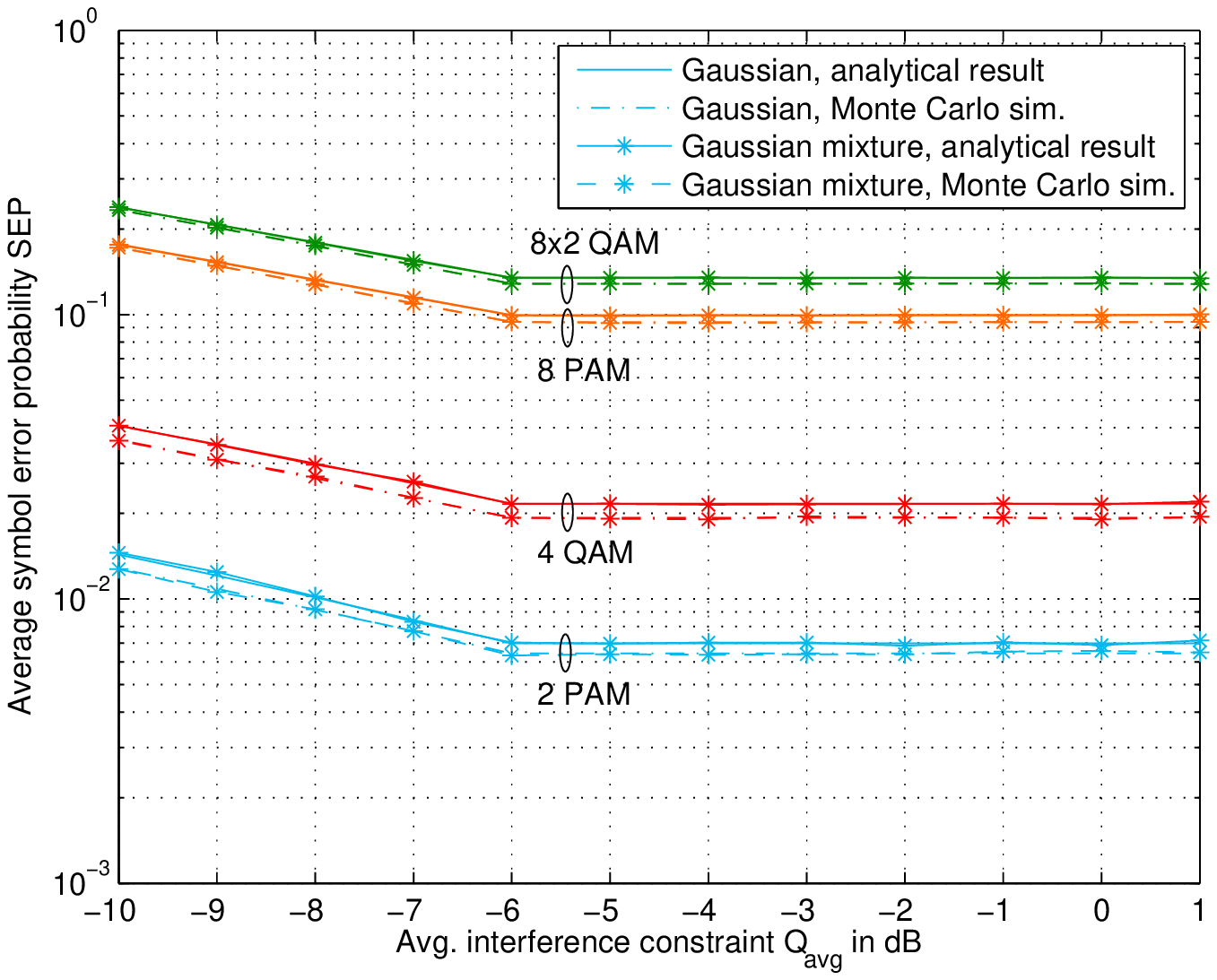}
\end{subfigure}
\begin{subfigure}[b]{0.45\textwidth}
\centering
\includegraphics[width=\textwidth]{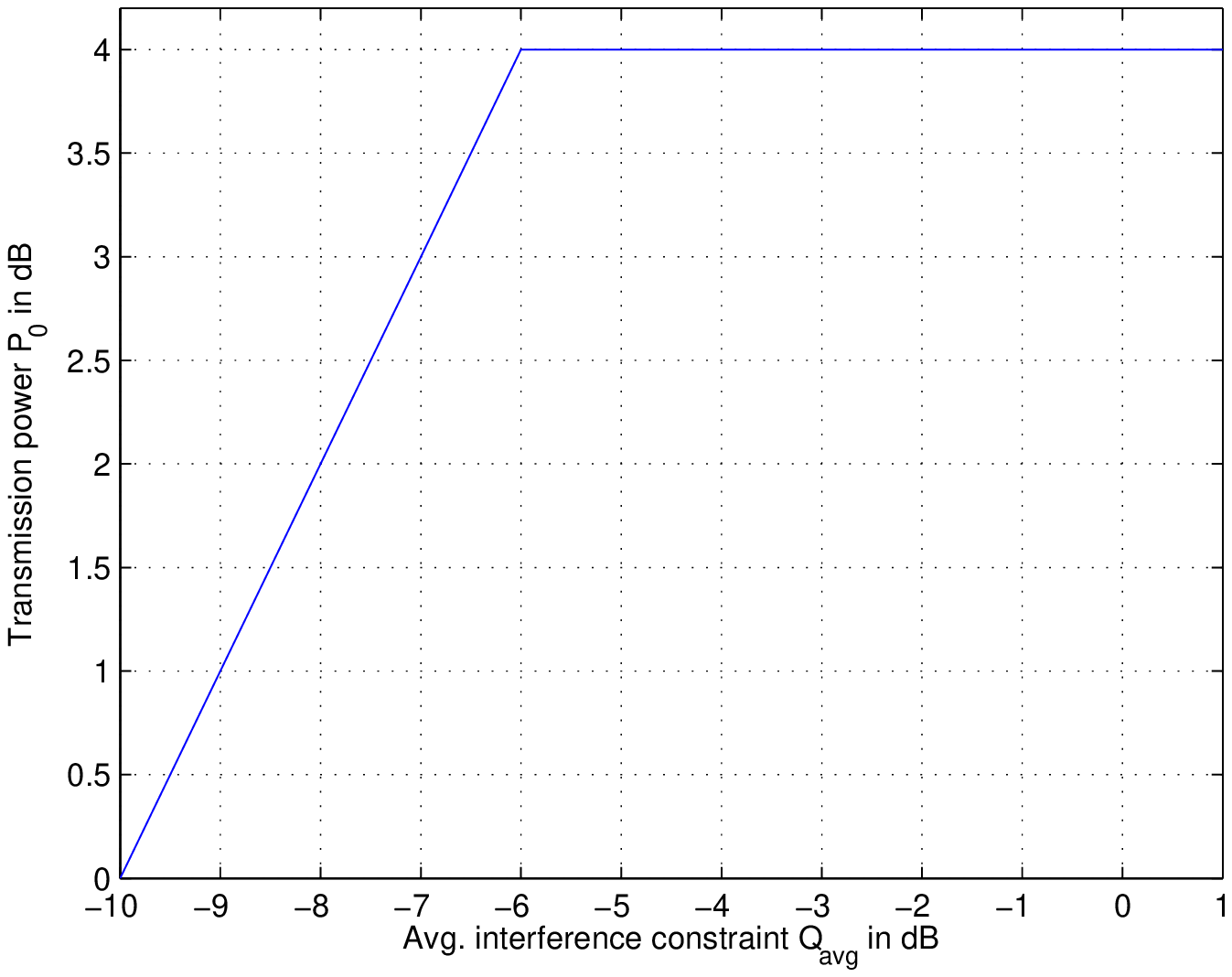}
\end{subfigure}
\caption{\small{Average symbol error probability $\SEP$ and transmission powers $P_0$ vs. average interference constraint, $Q_{\avg}$ in OSA scheme.}}\label{fig:SEP_P0_Qavg_osa}
\end{figure}

In Fig. \ref{fig:SEP_P0_Qavg_osa}, average $\SEP$ and transmission power $P_0$ are plotted as a function of the average interference constraint, $Q_{\avg}$, for the OSA scheme.  We again set $P_{\pk} = 4$ dB, $P_{\nid} =0.9$ and $P_{\f}=0.05$, and consider 2-PAM, 4-QAM, 8-PAM and $8\times 2$-QAM schemes. It is observed from the figure that as $Q_{\avg}$ increases, error probabilities initially decrease and then remain constant due to the fact that the secondary users can initially afford to transmit with higher transmission power $P_0$ as the interference constraint becomes less strict, but then get limited by the peak transmission power constraint and send data at the fixed power level of $P_{\pk}$. Again, we observe that lower error probabilities are attained when the primary user's received signal $w$ follows a Gaussian mixture distribution.

\begin{figure}
\centering
\begin{subfigure}[b]{0.45\textwidth}
\centering
\includegraphics[width=\textwidth]{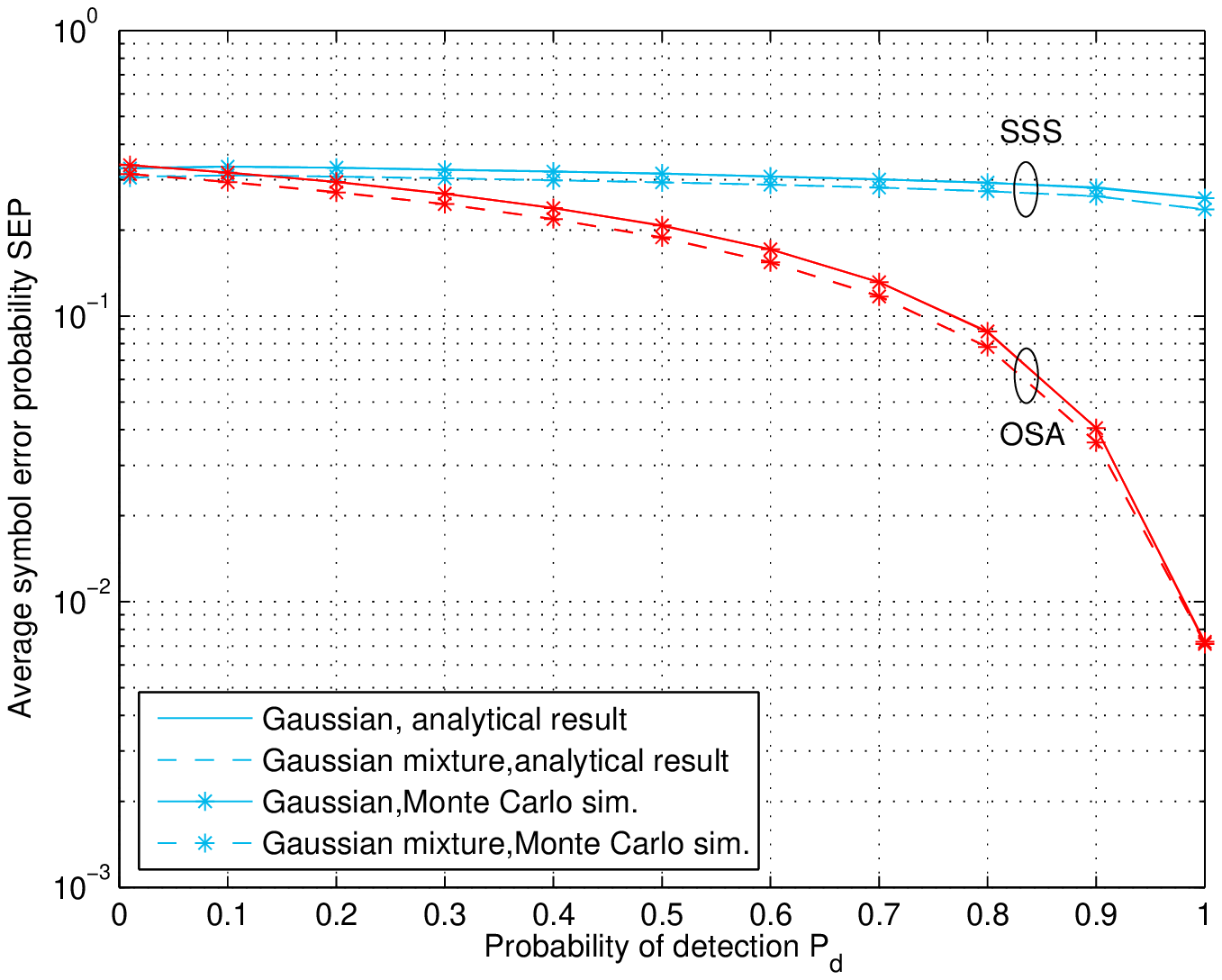}
\end{subfigure}
\begin{subfigure}[b]{0.45\textwidth}
\centering
\includegraphics[width=\textwidth]{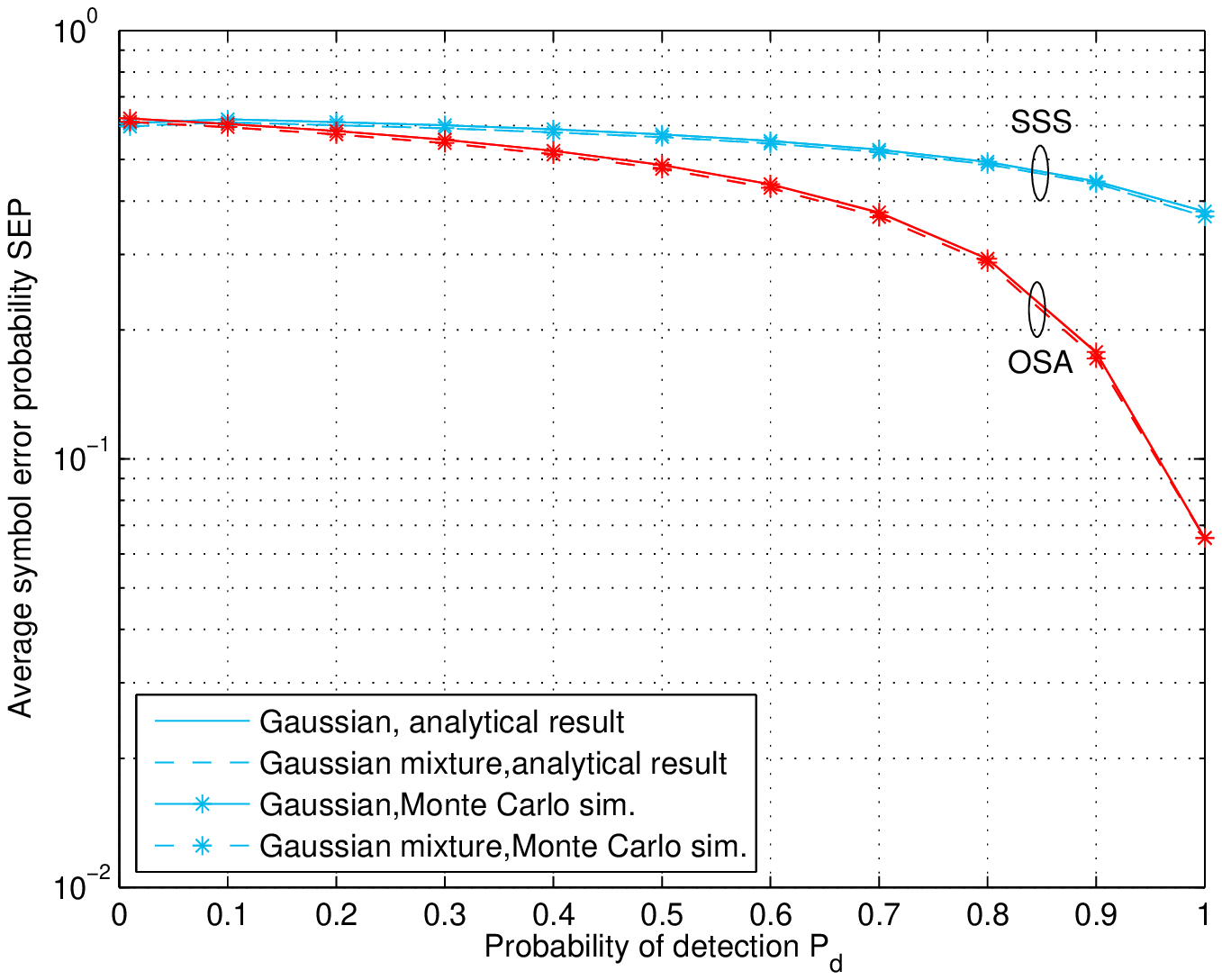}
\end{subfigure}
\caption{\small{Average symbol error probability $\SEP$ of 4-QAM (upper subfigure) and 8-PAM (lower subfigure) signaling vs. detection probability $P_{\nid}$ for SSS and OSA schemes.}}\label{fig:SEP_Pd_avg}
\end{figure}

In Fig. \ref{fig:SEP_Pd_avg}, the average $\SEP$s of 4-QAM (in the upper subfigure) and 8-PAM signaling (in the lower subfigure) are plotted as a function of the detection probability $P_{\nid}$. $P_{\f}$ is set to $0.05$. We consider both SSS and OSA schemes. It is assumed that $P_{\pk}=4$ dB and $Q_{\avg}=-10$ dB. We observe that $\SEP$ for both modulation types in both SSS and OSA schemes decreases as $P_{\nid}$ increases. Hence, performance improves with more reliable sensing. In this case, the primary reason is that more reliable detection enables the secondary users transmit with higher power in an idle-sensed channel. For instance, if $P_{\nid} = 1$, then the transmission power $P_0$  is only limited by $P_{\pk}$ in both SSS and OSA. In the figure, we also notice that lower $\SEP$ is achieved in the OSA scheme, when compared with the SSS scheme, due to the fact that OSA avoids transmission over a busy-channel in which interference from the primary user's transmission results in a more noisy channel and consequently higher error rates are experienced. At the same time, it is important to note that not transmitting in a busy-sensed channel as in OSA potentially reduces data rates.

\begin{figure}
\centering
\begin{subfigure}[b]{0.45\textwidth}
\centering
\includegraphics[width=\textwidth]{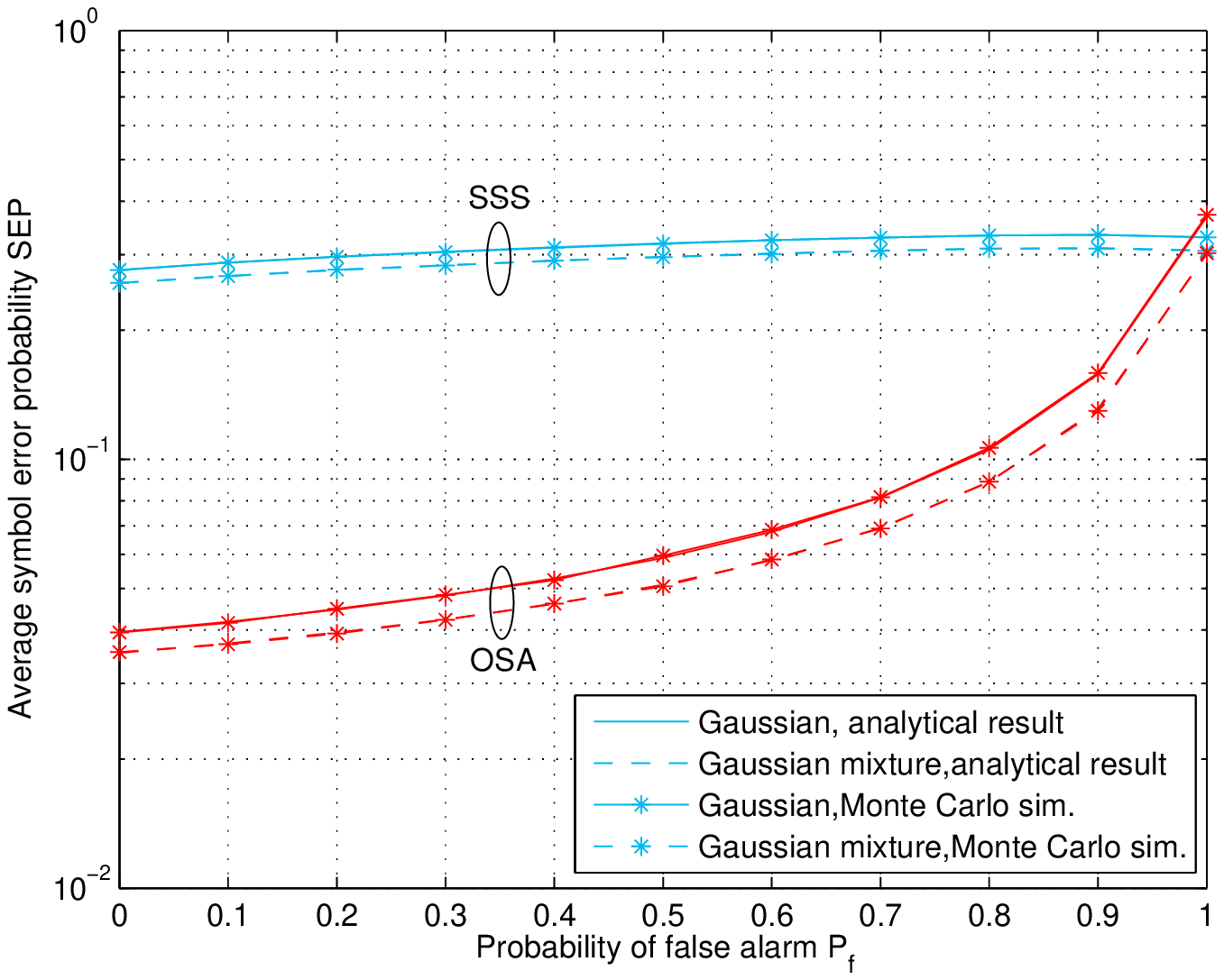}
\end{subfigure}
\begin{subfigure}[b]{0.4\textwidth}
\centering
\includegraphics[width=\textwidth]{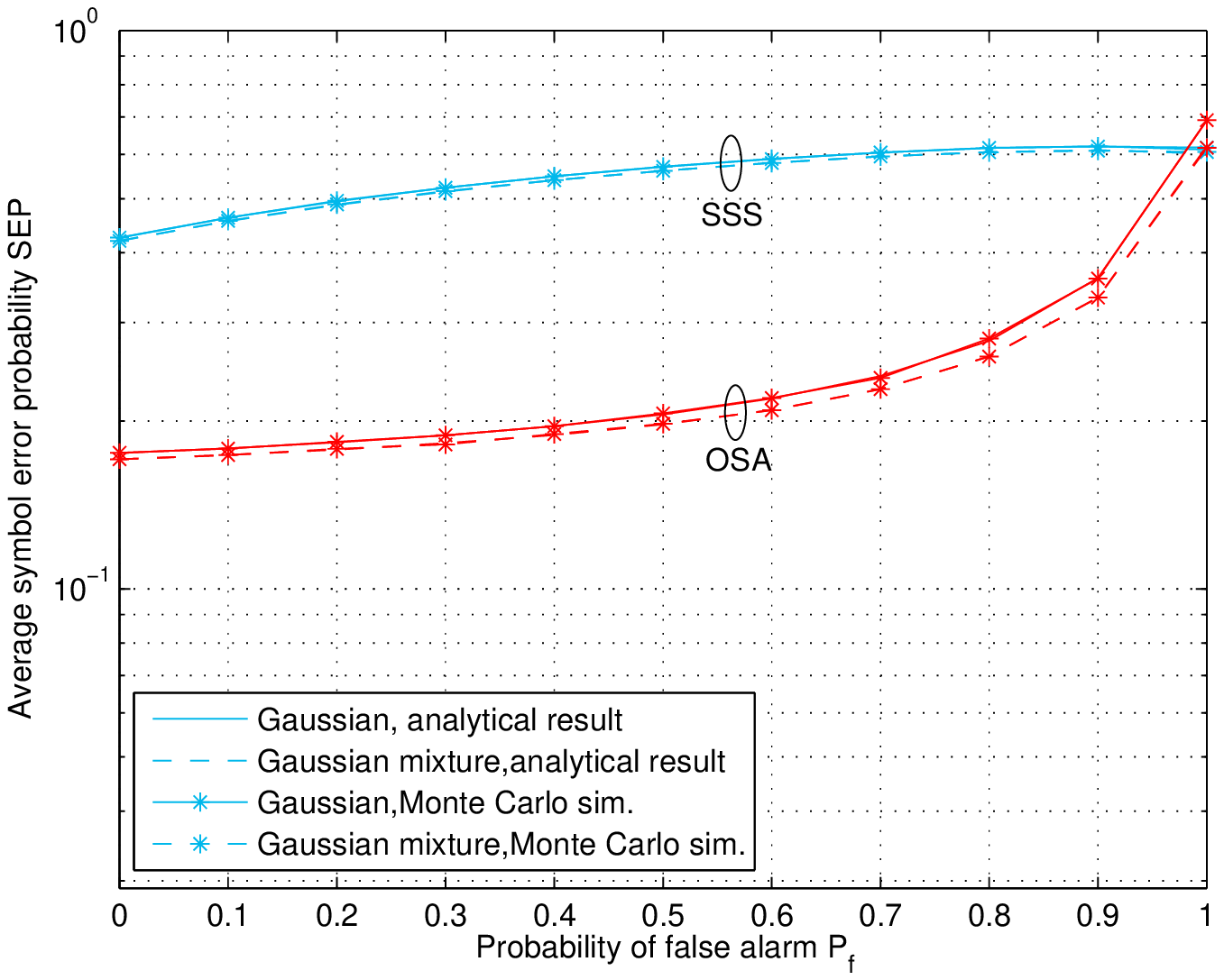}
\end{subfigure}
\caption{\small{Average symbol error probability $\SEP$ of 4-QAM (upper subfigure) and 8-PAM (lower subfigure) signaling vs. probability of false alarm $P_{\f}$ for SSS and OSA schemes.}}\label{fig:SEP_Pf_avg}
\end{figure}

In Fig. \ref{fig:SEP_Pf_avg}, the average $\SEP$s of 4-QAM and 8-PAM signaling are plotted as a function of the false alarm probability $P_{\f}$ for both SSS and OSA. It is assumed that $P_{\nid} = 0.9$. It is further assumed that $P_{\pk} = 4$ dB and $Q_{\avg} = -10$ dB, again corresponding to the case in which average interference power constraint is dominant compared to the peak transmit power constraint. In both schemes, $\SEP$ increases with increasing false alarm probability $P_{\f}$. Hence, degradation in sensing reliability leads to performance loss in terms of error probabilities. In OSA, the transmission power $P_0 = \min\left\{ P_{\pk}, \frac{Q_{\avg}}{(1-P_{\nid}) \E\{|g|^2\}}\right\}$ does not depend on $P_{\f}$ and hence is fixed in the figure. The increase in the error rates can be attributed to the fact that secondary users more frequently experience interference from primary user's transmissions due to sensing uncertainty. For instance, in the extreme case in which $P_{\f} = 1$, the probability terms in (\ref{eq:Pe_OSA_without_power_constraints}) become $\Pr\{\mH_0|\hH_0\} = 0$ and $\Pr\{\mH_1|\hH_0\} = 1$, indicating that although the channel is sensed as idle, it is actually busy with probability one and the additive disturbance in OSA transmissions always includes $w$. In the SSS scheme, higher $P_{\f}$ leads to more frequent transmissions with power $P_1$ which is generally smaller than $P_0$ in order to limit the interference on the primary users. Transmission with smaller power expectedly increases the error probabilities. On the other hand, we interestingly note that as $P_{\f}$ approaches 1, $P_1$ becomes higher than $P_0$ when (\ref{eq:SSS_Expectation_Pe}) is solved, resulting in a slight decrease in $\SEP$ when $P_{\f}$ exceeds around 0.9.

\subsection{SEP under Peak Interference Constraints}

We now address the peak interference constraints by assuming that the transmission powers are limited as in (\ref{eq:instantaneous-power-g}). In this section, analytical error probability curves are plotted using the upper bounds in (\ref{eq:Pe_SSS_min}) in the case of SSS and in (\ref{eq:Pe_OSA_min}) in the case of OSA since we only have closed-form expressions for the error probability upper bounds when we need to evaluate an additional expectation with respect to $|g|^2$. Note that these upper bounds provide exact error probabilities when PAM is considered. Additionally, the discrepancy in QAM is generally small as demonstrated through comparisons with Monte Carlo simulations which provide the exact error rates in the figures.

\begin{figure}[h]
\centering
\begin{subfigure}[b]{0.45\textwidth}
\centering
\includegraphics[width=\textwidth]{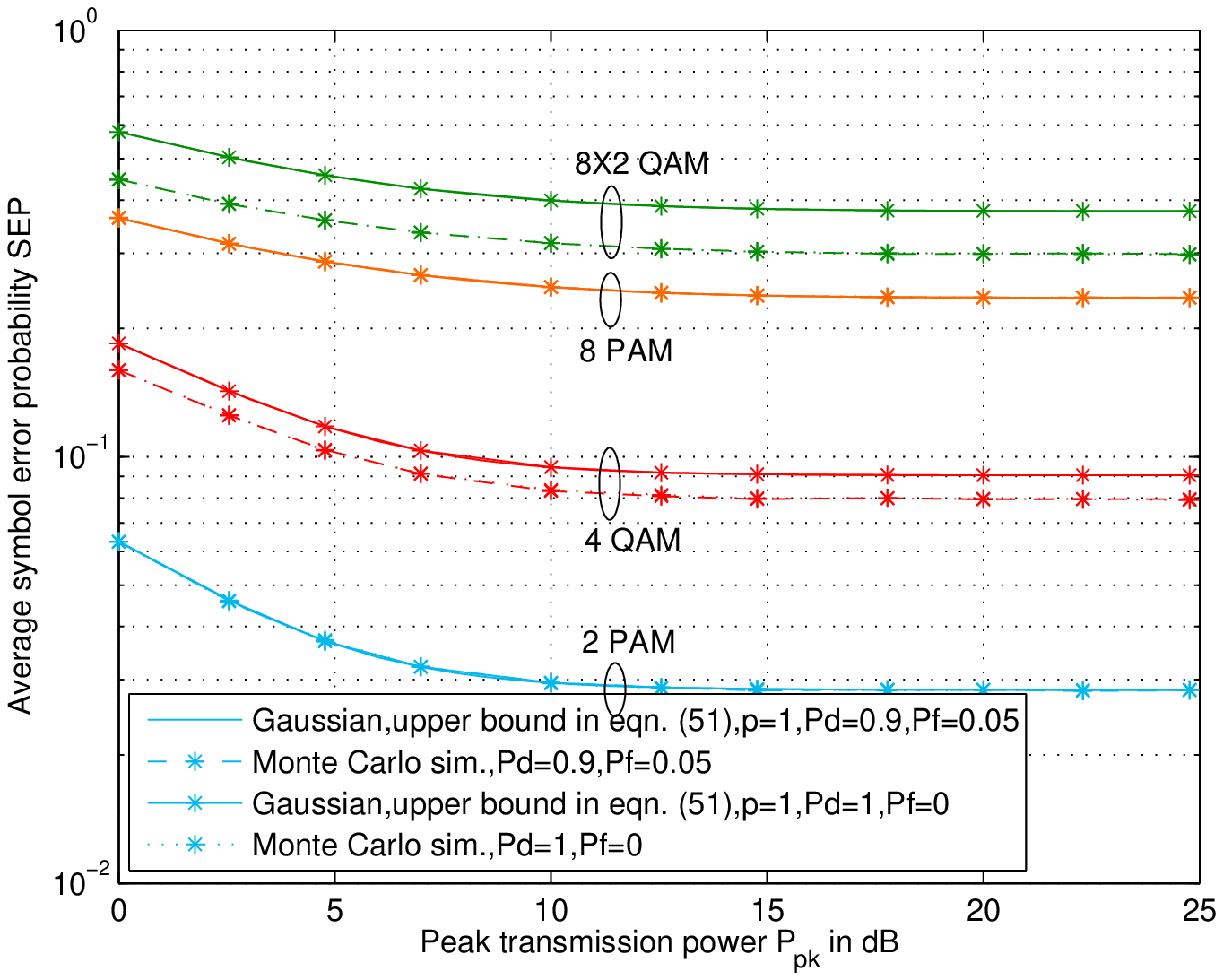}
\end{subfigure}
\begin{subfigure}[b]{0.45\textwidth}
\centering
\includegraphics[width=\textwidth]{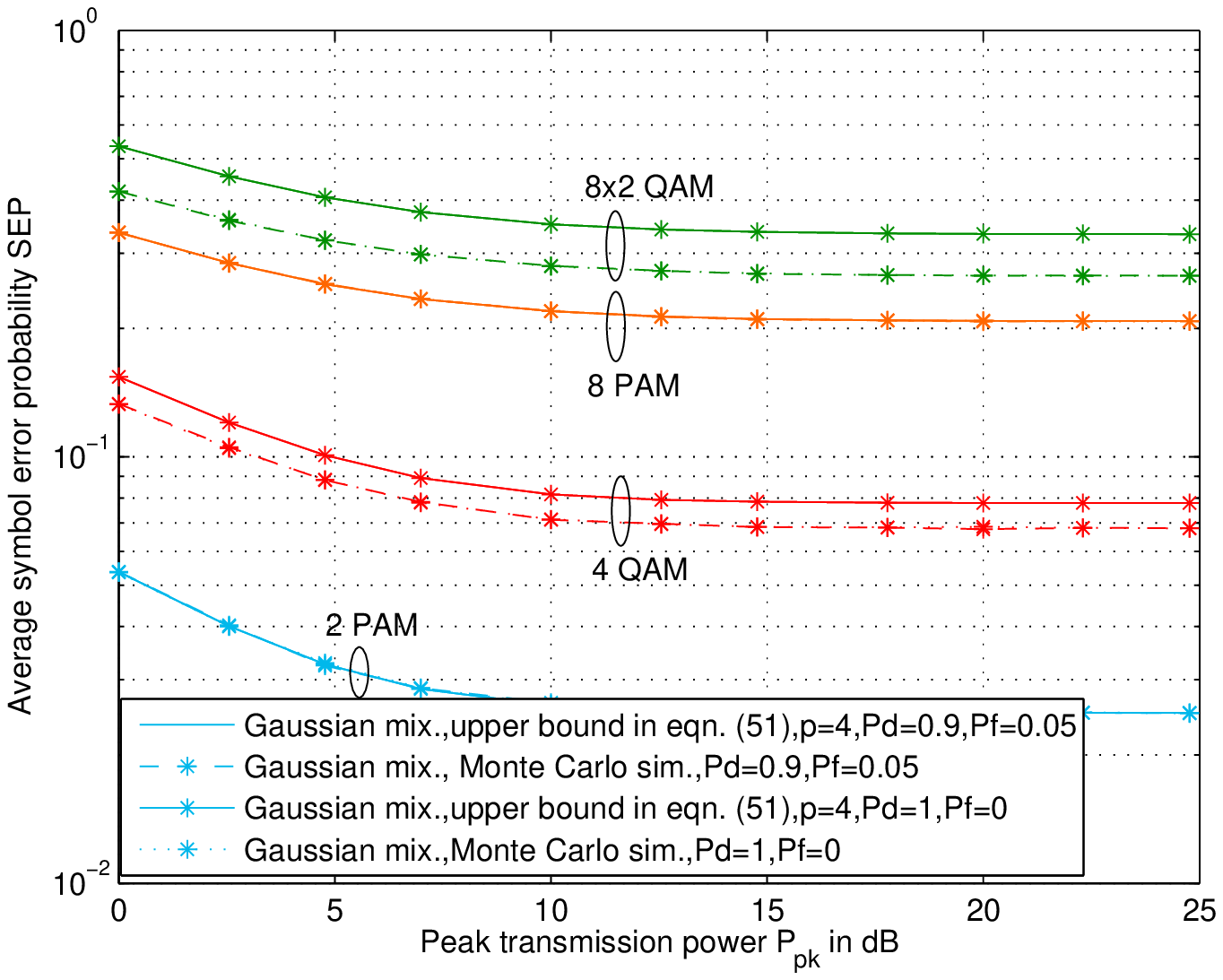}
\end{subfigure}
\caption{\small{Average symbol error probability $\SEP$ vs. peak transmission power $P_{\pk}$ in dB for SSS scheme when the primary user signal is modeled by Gaussian distribution (upper subfigure) and Gaussian mixture distribution (lower subfigure).}}\label{fig:SEP_Ppk_SSS}
\end{figure}

In Fig. \ref{fig:SEP_Ppk_SSS}, we plot the average $\SEP$ as a function of the peak transmission power, $P_{\pk}$, for the SSS scheme in the presence of Gaussian distributed and Gaussian-mixture distributed primary user's received faded signal $w$ in the upper and lower subfigures, respectively. The secondary users again employ 2-PAM, 4-QAM, 8-PAM and $8\times 2$-QAM schemes. The peak interference power constraint, $Q_{\pk}$ is set to 4 dB. It is seen that Monte Carlo simulations match with the analytical results for PAM and are slightly lower than the analytical upper bounds for QAM. As expected, the average $\SEP$ initially decreases with increasing $P_{\pk}$ and a higher modulation size leads to higher error rates at the same transmission power level. We again observe that lower error rates are experienced when $w$ has a Gaussian mixture distribution rather than a Gaussian distribution with the same variance. It is also seen that as $P_{\pk}$ increases, the $\SEP$ curves in all cases approach some error floor at which point interference constraints become the limiting factor.

Another interesting observation is the following. In Fig. \ref{fig:SEP_Ppk_SSS}, $\SEP$s are plotted for two different pairs of detection and false alarm probabilities. In the first scenario, channel sensing is perfect, i.e., $P_{\nid} = 1$ and $P_{\f} = 0$. In the second scenario, we have $P_{\nid} = 0.9$ and $P_{\f} = 0.05$. In both scenarios, we observe the same error rate performance. This is because the same transmission power is used regardless of whether the channel is detected as idle or busy, i.e., $P_i^*=\min\left\{P_{\pk},\frac{Q_{\pk}}{|g|^2}\right\}$ for both $i = 0,1$. The interference constraints are very strict as noted in Section \ref{subsec:power-interference}. Hence, averaging over channel sensing decisions becomes averaging over the prior probabilities of channel occupancy, which does not depend on the probabilities of detection and false alarm. Indeed, spectrum sensing can be altogether omitted under these constraints.

\begin{figure}[h]
\centering
\begin{subfigure}[b]{0.45\textwidth}
\centering
\includegraphics[width=\textwidth]{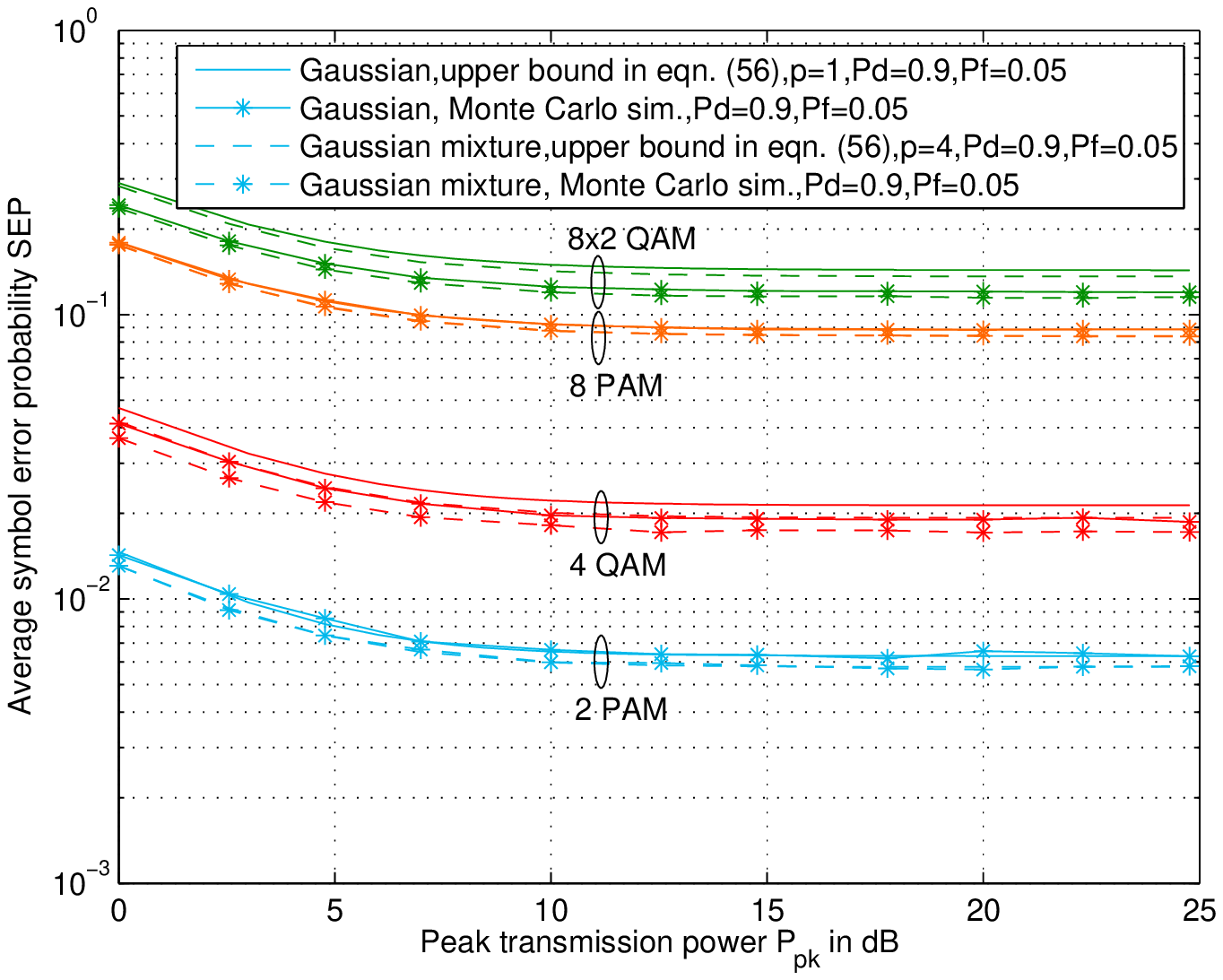}
\end{subfigure}
\begin{subfigure}[b]{0.45\textwidth}
\centering
\includegraphics[width=\textwidth]{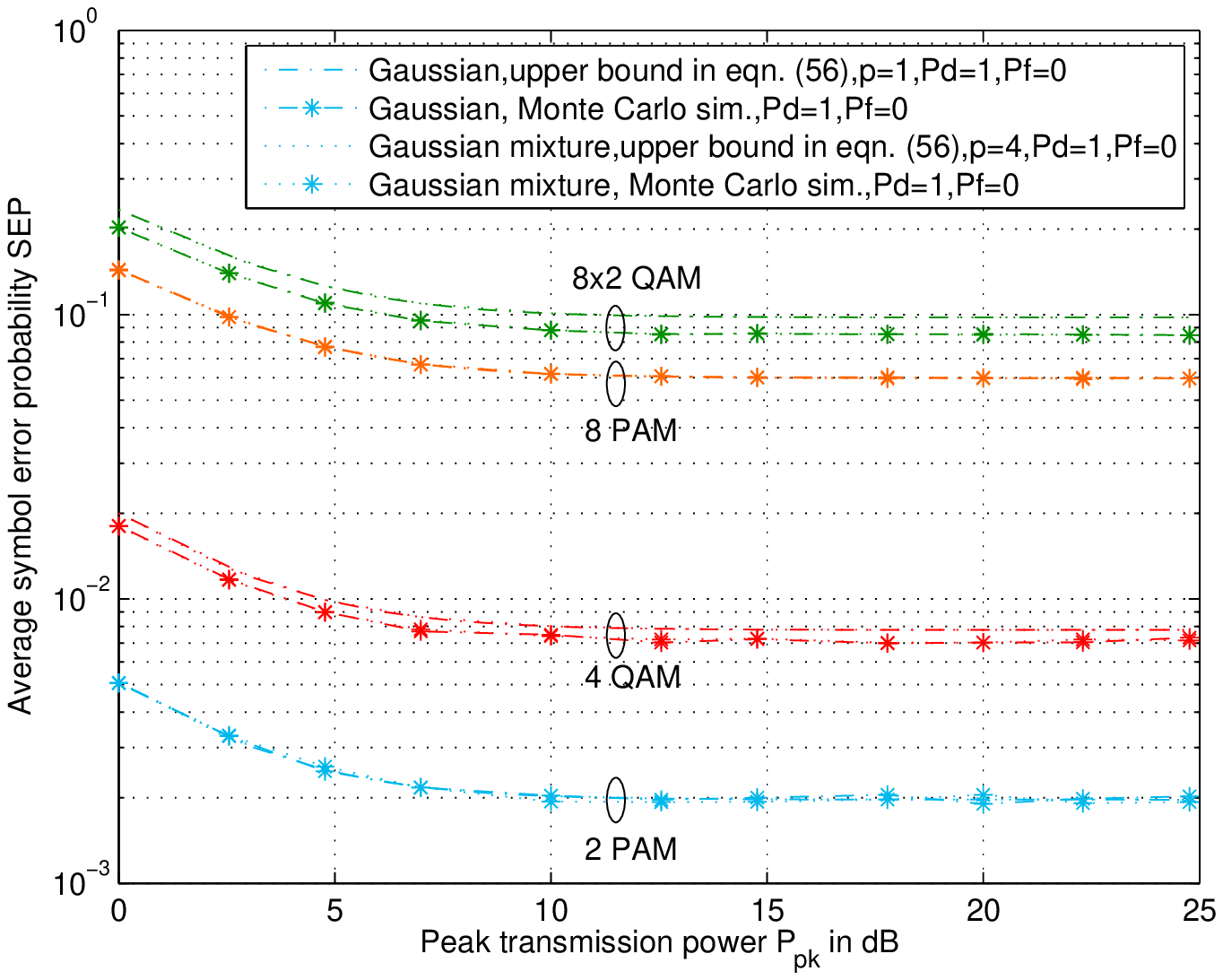}
\end{subfigure}
\caption{\small{Average symbol error probability $\SEP$ vs. peak transmission power $P_{\pk}$ in dB for OSA scheme in the presence of Gaussian and Gaussian mixture primary user's interference signal under imperfect sensing result (upper subfigure) and perfect sensing result (lower subfigure).}}
\label{fig:SEP_Ppk_OSA}
\end{figure}

In Fig. \ref{fig:SEP_Ppk_OSA}, we plot the average $\SEP$ as a function of $P_{\pk}$ for the OSA scheme. As before, 2-PAM, 4-QAM, 8-PAM and $8\times 2$-QAM are considered. Imperfect sensing with $P_{\nid} = 0.9$ and $P_{\f} = 0.05$ is considered in the upper subfigure whereas perfect sensing (i.e., $P_{\nid} = 1$ and $P_{\f} = 0$) is assumed in the lower subfigure. In both subfigures, it is seen that increasing $P_{\pk}$ initially reduces $\SEP$ which then hits an error floor as the interference constraints start to dominate. It is also observed that perfect channel sensing improves the error rate performance of cognitive users. Note that if sensing is perfect, secondary users transmit only if the channel is actually idle and experience only the background noise $n$. On the other hand, under imperfect sensing, secondary users transmit in miss-detection scenarios as well, in which they are affected by both the background noise and primary user interference $w$, leading to higher error rates. Cognitive radio transmission impaired by Gaussian mixture distributed $w$ again results in lower $\SEP$ compared to Gaussian distributed $w$. But, of course, this distinction disappears with perfect sensing in the lower subfigure since the secondary users experience only the Gaussian background noise $n$ as noted above. Finally, the gap between the analytical and simulation results for QAM is due to the use of upper bounds in the analytical error curves as discussed before.


\begin{figure}[h]
\centering
\begin{subfigure}[b]{0.45\textwidth}
\centering
\includegraphics[width=\textwidth]{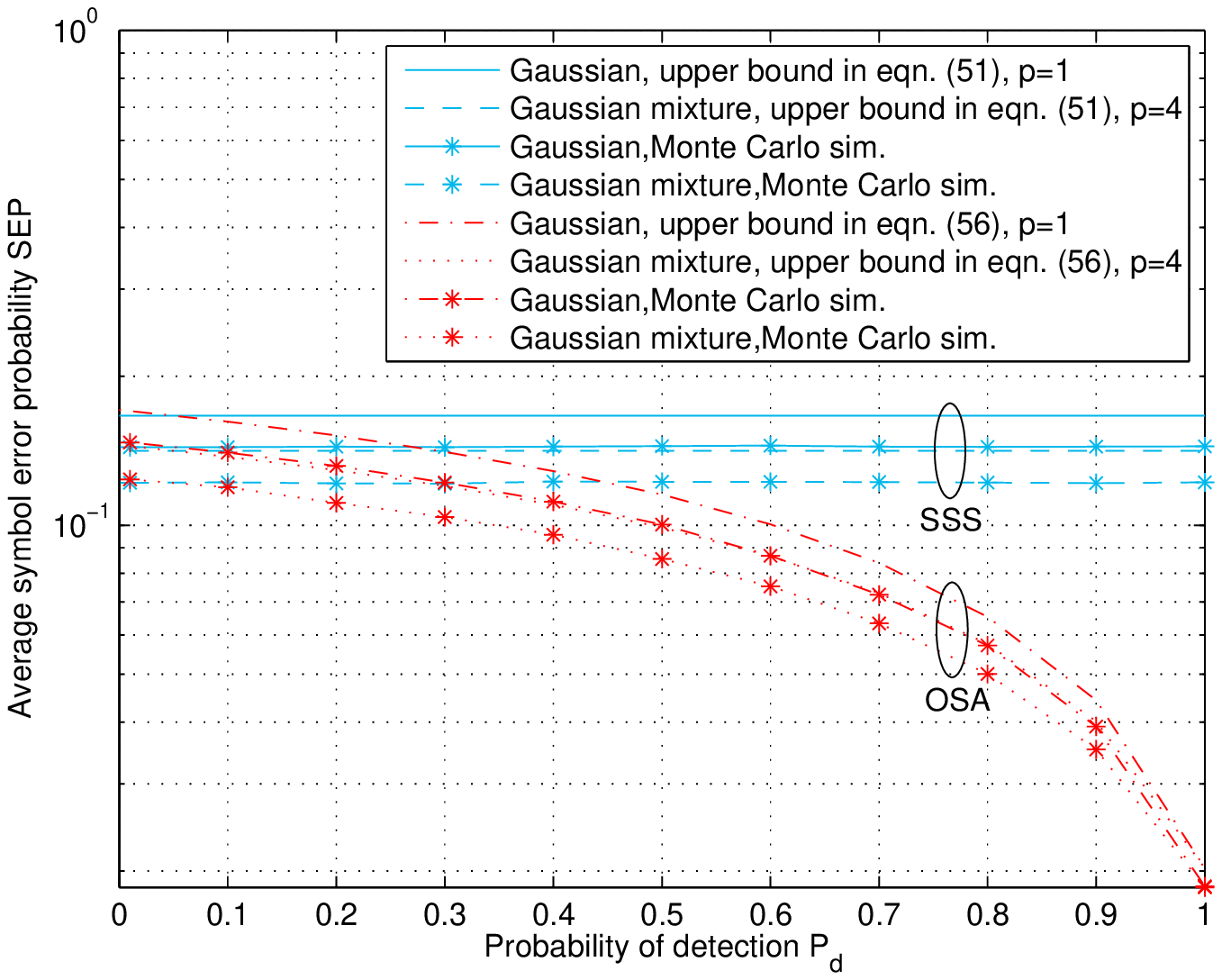}
\end{subfigure}
\begin{subfigure}[b]{0.45\textwidth}
\centering
\includegraphics[width=\textwidth]{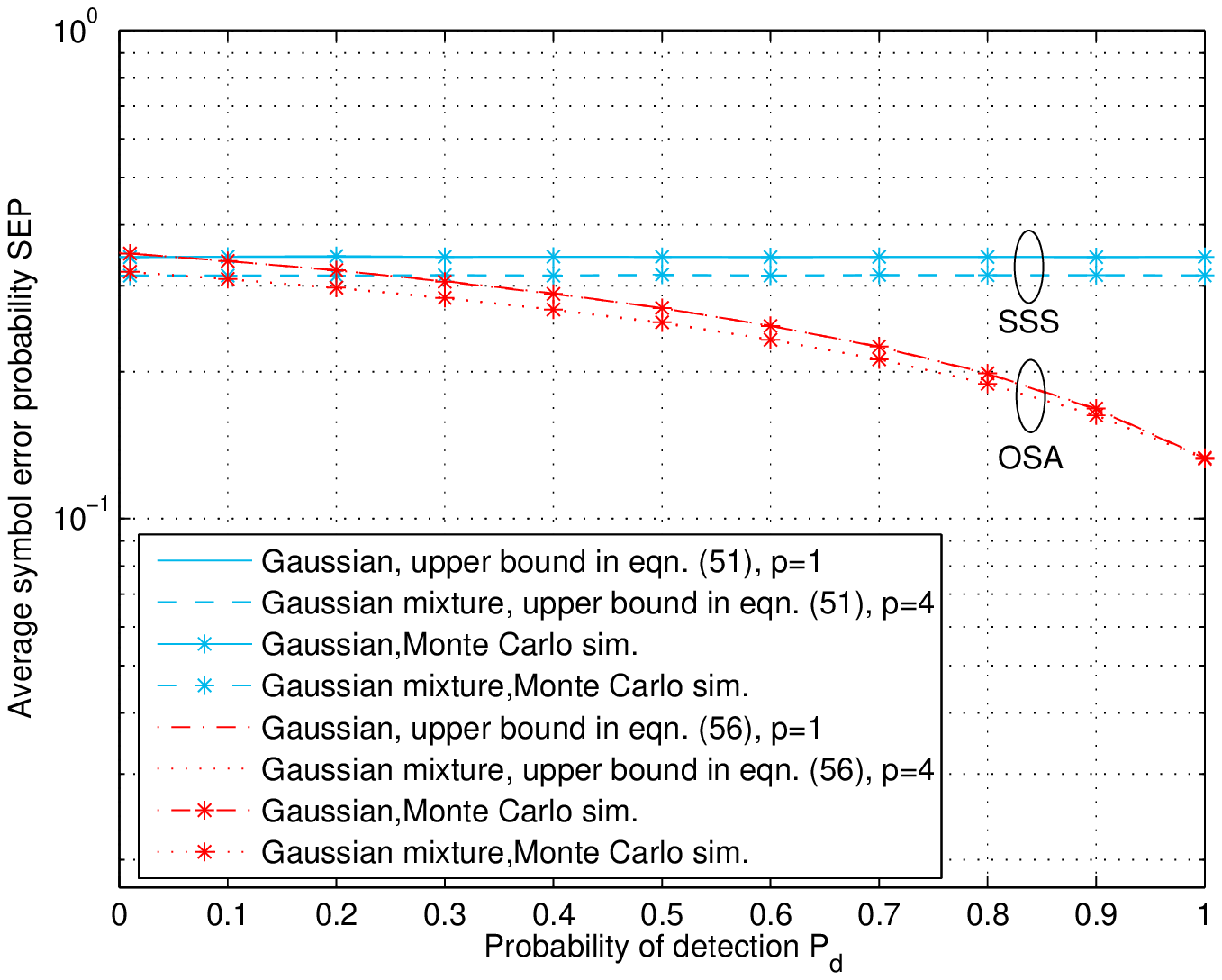}
\end{subfigure}
\caption{\small{Average symbol error probability $\SEP$ of 4-QAM (upper subfigure) and 8-PAM (lower subfigure) signaling vs. detection probability $P_{\nid}$ for SSS and OSA schemes.}}
\label{fig:SEP_Pd_Qpk}
\end{figure}

In Fig. \ref{fig:SEP_Pd_Qpk}, we display the average $\SEP$ of 4-QAM and 8-PAM signaling as a function of the detection probability $P_{\nid}$. $P_{\f}$ is set to $0.05$. Both SSS and OSA schemes are considered. Here, we also assume that $P_{\pk}= 4$ dB, $Q_{\pk}=0$ dB. It is seen that error rate performances for SSS scheme for both modulation types do not depend on detection probability because of the same reasoning explained in the discussion of Fig. \ref{fig:SEP_Ppk_SSS}. On the other hand, the error rate performance for the OSA scheme improves with increasing detection probability since the secondary user experiences less interference from the primary user activity. It is also seen that OSA scheme outperforms SSS scheme.

\begin{figure}[h]
\centering
\begin{subfigure}[b]{0.45\textwidth}
\centering
\includegraphics[width=\textwidth]{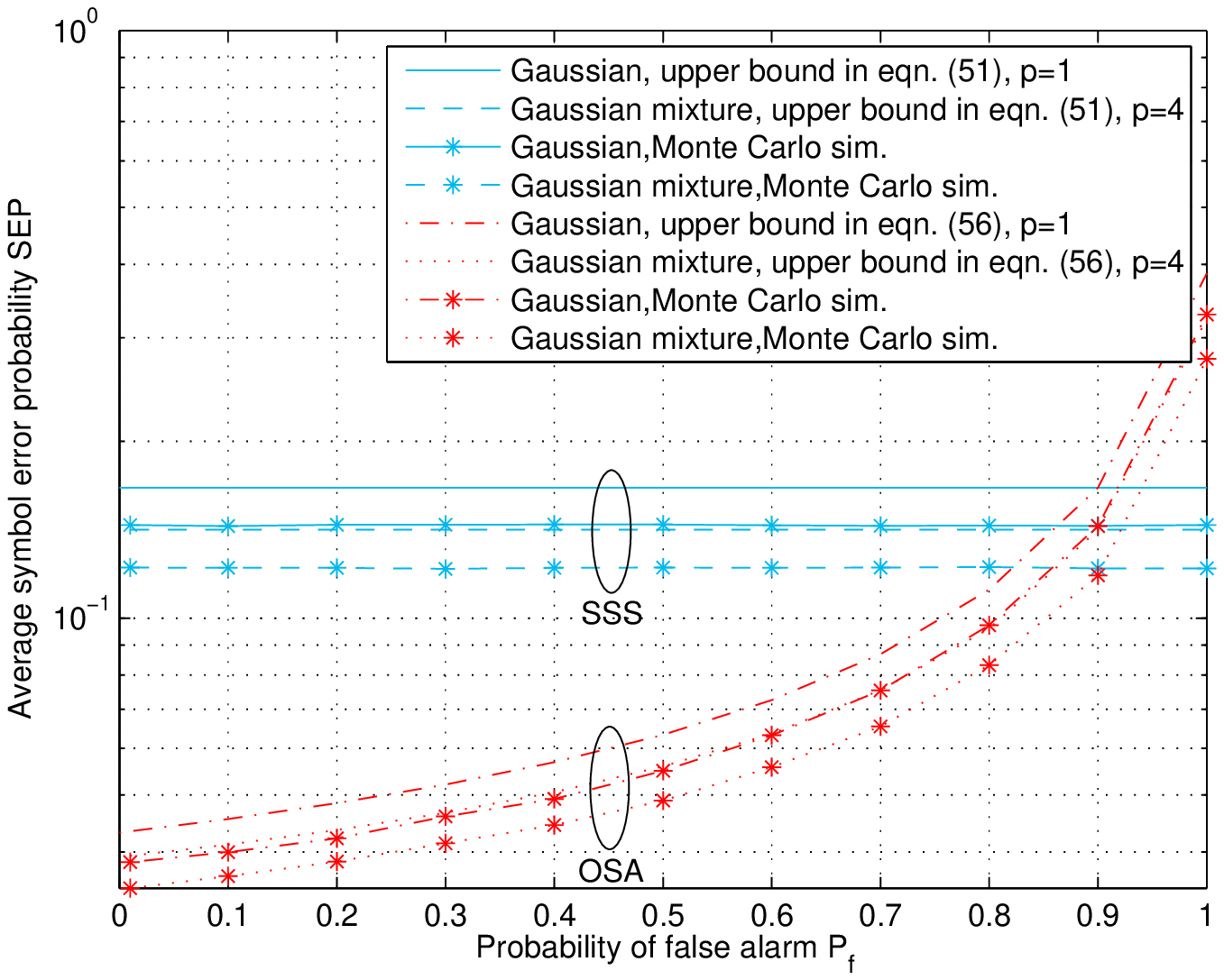}
\end{subfigure}
\begin{subfigure}[b]{0.45\textwidth}
\centering
\includegraphics[width=\textwidth]{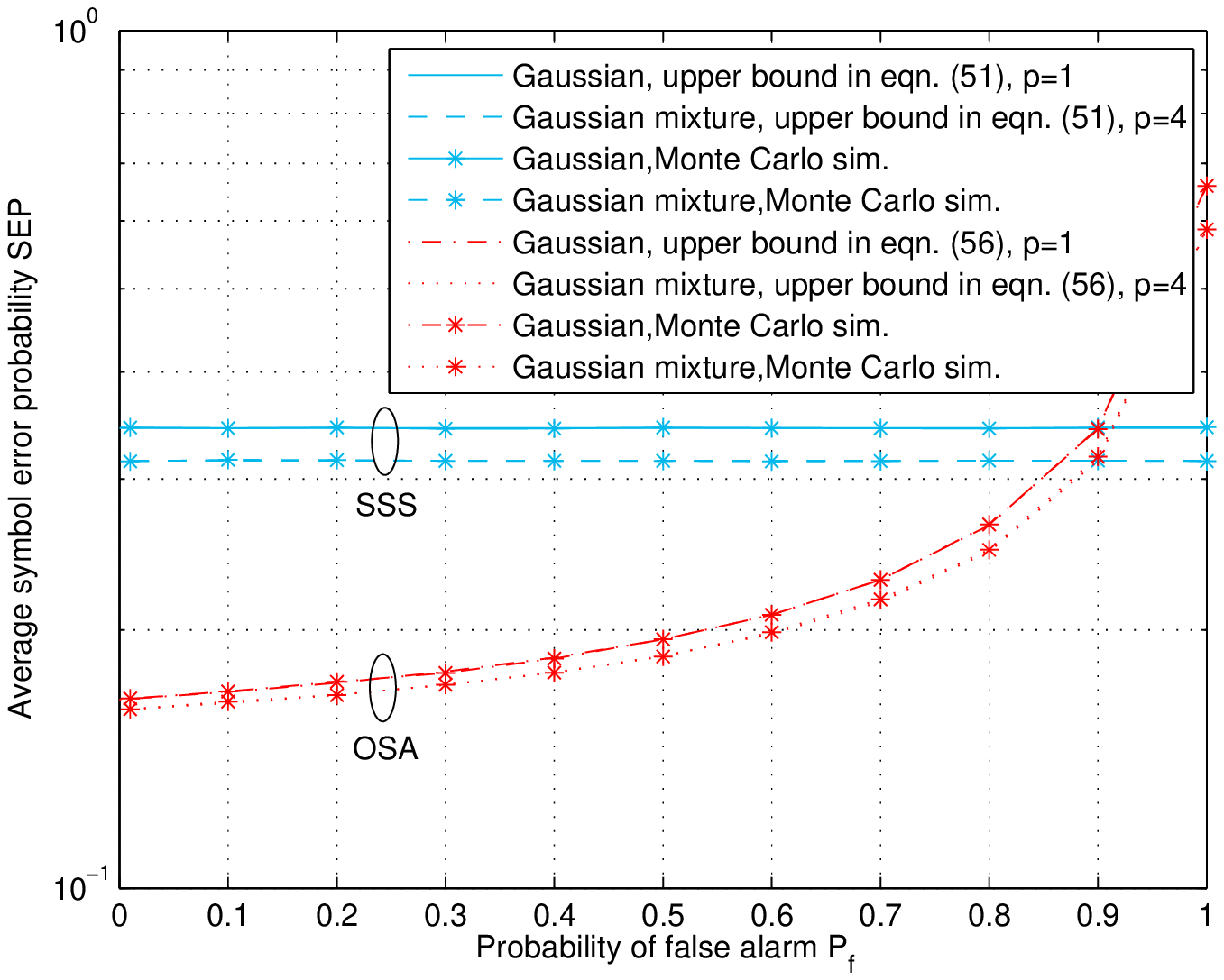}
\end{subfigure}
\caption{\small{Average symbol error probability $\SEP$ of 4-QAM (upper subfigure) and 8-PAM (lower subfigure) signaling vs. probability of false alarm $P_{\f}$ for SSS and OSA schemes.}}
\label{fig:SEP_Pf_Qpk}
\end{figure}

In Fig. \ref{fig:SEP_Pf_Qpk}, we analyze the average $\SEP$ of 4-QAM and 8-PAM signaling as a function of the false alarm probability $P_{\f}$. Detection probability is $P_{\nid} = 0.9$. Similarly as before, $P_{\pk} = 4$ dB and $Q_{\pk} = 0$ dB. Again, error rate performance does not depend on $P_{\f}$ in the SSS scheme. It is observed that $\SEP$ in OSA scheme increases with increasing false alarm probability. Hence, degradation in the sensing performance in terms of increased false alarm probabilities leads to degradation in the error performance. As discussed in Section \ref{subsec:averageinterference} regarding the error rates in Fig. \ref{fig:SEP_Pf_avg}, deterioration in the performance is due to more frequent exposure to interference from primary user's transmissions in the form of $w$.
One additional remark from the figure is that SSS scheme gives better error rate performance compared to OSA scheme at higher values of $P_{\f}$.

\section{Conclusion}\label{sec:conclusion}

We have studied the error rate performance of cognitive radio transmissions in both SSS and OSA schemes in the presence of transmit and interference power constraints, sensing uncertainty, and Gaussian mixture distributed interference from primary user transmissions. In this setting, we have proved that the midpoints between the signals are optimal thresholds for the detection of equiprobable rectangular QAM signals. We have first obtained exact $\SEP$ expressions for given fading realizations and then derived closed-form average SEP expressions for the Rayleigh fading channel. We have further provided upper bounds on the error probabilities averaged over the fading between the secondary transmitter and primary receiver under the peak interference constraint. The analytical $\SEP$ expressions have been validated through Monte-Carlo simulations.

In the numerical results, we have had several interesting observations. We have seen that, when compared to SSS, lower error rates are generally attained in the OSA scheme. Also, better error performance is achieved in the presence of Gaussian-mixture distributed $w$ in comparison to that achieved when $w$ is Gaussian distributed with the same variance. We have also addressed how the error rates and transmission powers vary as a function of the power and interference constraints. Finally, we have demonstrated that symbol error probabilities are in general dependent on sensing performance through the detection and false alarm probabilities. For instance, we have observed that as the detection probability increases, the error rate performance under both schemes improves in interference-limited environments. Similarly, SEP is shown to decrease with decreasing false-alarm probability. Hence, we conclude that sensing performance is tightly linked to error performance and improved sensing leads to lower error rates.

\appendix

\subsection{Proof of Proposition \ref{prop:decisionrule}} \label{app:proof-Prop}

Since the signals are \emph{equiprobable}, the maximum likelihood (ML) decision rule is optimal in the sense that it minimizes the average probability of error \cite{Poor}. Since cognitive transmission is allowed only when the channel is sensed as idle under OSA scheme, it is enough to evaluate the ML decision rule under sensing decision $\hH_0$, which can be expressed as
\begin{align}
\hat{m} = \argmax_{0\le m\le M-1} f(\by | s_{n,q}, h, \hH_0).
\end{align}
The above decision rule can further be expressed as
\begin{equation}
\begin{split}
\hat{m} = \argmax_{0\le m\le M-1} \Big(&\Pr\{\mH_0|\hH_{0}\}f(\by | s_{n,q}, h, \hH_{0}, \mH_0) \\+ &\Pr\{\mH_1|\hH_{0}\}f(\by | s_{n,q}, h, \hH_{0}, \mH_1)\Big).
\end{split}
\end{equation}
Above maximization simply becomes the comparison of the likelihood functions of the received signals given the transmitted signals $s_{n,q}$. Without loss of generality, we consider the signal constellation point $s_{n,q}$. Then, the decision region for the in-phase component of this signal constellation point is given by (\ref{eq:comparision_ml1}).
\begin{figure*}
\begin{align}
\label{eq:comparision_ml1}
&\sum_{j=0}^{1}\Pr\{\mH_j | \hH_{0}\}f(\by | s_{n,q}, h, \hH_{0}, \mH_j) \geq \sum_{j=0}^{1}\Pr\{\mH_j | \hH_{0}\}f(\by | s_{n+1,q}, h, \hH_{0}, \mH_j) \\
\label{eq:comparision_ml2}
&\sum_{j=0}^{1}\Pr\{\mH_j | \hH_{0}\}f(\by | s_{n,q}, h, \hH_{0}, \mH_j) \geq \sum_{j=0}^{1}\Pr\{\mH_j | \hH_{0}\}f(\by | s_{n-1,q}, h, \hH_{0}, \mH_j)
\end{align}
\hrule
\end{figure*}
The right-side boundary of the corresponding decision region, which can be found by equating the likelihood functions in (\ref{eq:comparision_ml1})--(\ref{eq:comparision_ml2}) shown at the top of next page. Gathering the common terms together, the expression in (\ref{eq:comparision_ml1}) can further be written as in (\ref{eq:comparision_ml3}) given on the next page.
\begin{figure*}
\begin{align}
\small
&\frac{\Pr\{\mH_0|\hH_{0}\}}{2\pi\sigma_{n}^2}\rme^{-\frac{( \by_r - s_n|h|)^2+(\by_i - s_q|h|)^2}{2\sigma_{n}^2}}
\left(\!1-\rme^{\frac{2d_{min,0}|h|( \by_r - s_n|h|)-d_{min,0}^2|h|^2}{2\sigma_{n}^2}}\!\right)
\nonumber \\
&+ \Pr\{\mH_1 |\hH_{0}\}\sum_{l=1}^p\frac{\lambda_l}{2\pi(\sigma_{l}^2 + \sigma_{n}^2)}\rme^{-\frac{( \by_r - s_n|h|)^2+(\by_i - s_q|h|)^2}{2(\sigma_{l}^2 + \sigma_{n}^2)}}\left(\!1 - \rme^{\frac{2d_{min,0}|h|( \by_r - s_n|h|)-d_{min,0}^2|h|^2}{2(\sigma_{l}^2 + \sigma_{n}^2)}}\!\right) \geq 0. \label{eq:comparision_ml3}
\end{align}
\normalsize
\hrule
\end{figure*}

We note that all the terms on the left-hand side of (\ref{eq:comparision_ml3}) other than the terms inside the parentheses are nonnegative. Let us now consider these difference terms. Inside the first set of parentheses, we have
\begin{align}
\label{eq:exponential_compare1}
1 - \rme^{\big(\frac{2(\by_r - s_n|h|) d_{min,0}|h|-d_{min,0}^2|h|^2}{2\sigma_{n}^2}\big)}
\end{align}
which can easily be seen to be greater than zero if $\by_r - s_n|h| < \frac{d_{min,0}|h|}{2}$ and is zero if $\by_r - s_n|h| = \frac{d_{min,0}|h|}{2}$. The same is also true for the term inside the second set of parentheses given by
\begin{gather}
1 - \rme^{\big(\frac{2(\by_r - s_n|h|) d_{min,0}|h|- d_{min,0}^2|h|^2}{2(\sigma_{l}^2 + \sigma_{n}^2)}\big)}.
\end{gather}
Therefore, the inequality in (\ref{eq:comparision_ml3}) can be reduced to
\begin{gather}
\by_r - s_n|h| \le \frac{d_{min,0}|h|}{2}. \label{eq:comparision_ml3-simplified}
\end{gather}
Similarly, (\ref{eq:comparision_ml2}) simplifies to
\begin{gather}
\by_r - s_n|h| \ge -\frac{d_{min,0}|h|}{2}. \label{eq:comparision_ml2-simplified}
\end{gather}
From these observations, we immediately conclude that the decision rule to detect $s_n$ involves comparing $\by_r$ with thresholds located at midpoints between the received neighboring signals.

Following the same approach, we can determine the decision region for the quadrature component of the signal constellation point $s_{n,q}$ by comparing the likelihood functions in (\ref{eq:comparision_ml1_im}) -- (\ref{eq:comparision_ml2_im}) shown on the next page,
\begin{figure*}
\begin{align}
\label{eq:comparision_ml1_im}
&\sum_{j=0}^{1}\Pr\{\mH_j | \hH_{0}\}f(\by | s_{n,q}, h, \hH_{0}, \mH_j) \geq \sum_{j=0}^{1}\Pr\{\mH_j | \hH_{0}\}f(\by | s_{n,q+1}, h, \hH_{0}, \mH_j) \\
\label{eq:comparision_ml2_im}
&\sum_{j=0}^{1}\Pr\{\mH_j | \hH_{0}\}f(\by | s_{n,q}, h, \hH_{0}, \mH_j) \geq \sum_{j=0}^{1}\Pr\{\mH_j | \hH_{0}\}f(\by | s_{n,q-1}, h, \hH_{0}, \mH_j).
\end{align}
\hrule
\end{figure*}
which similarly reduce to $\by_i - s_q|h| \le \frac{d_{min,0}|h|}{2}$ and $\by_i - s_q|h| \ge -\frac{d_{min,0}|h|}{2}$, respectively. Hence, we again have the thresholds at midpoints between the neighboring received signals.

Under the SSS scheme, the secondary users are allowed to transmit in busy-sensed channel (i.e., under sensing decision $\hH_1$) as well. Since the simplified decision rules in (\ref{eq:comparision_ml3-simplified}) and (\ref{eq:comparision_ml2-simplified}) do not depend on the sensing decision $\hH_i$, the same set of inequalities are obtained for the ML detection rule under sensing decision $\hH_1$, leading to the same conclusion regarding the decision rule and thresholds. \hfill $\square$


\end{document}